\documentclass[preprint,12pt]{elsarticle}

\usepackage{amsmath}
\usepackage{amssymb}
\usepackage{caption}
\usepackage{graphicx}
\usepackage{latexsym}
\usepackage{times}
\usepackage[version=4]{mhchem}
\usepackage{lineno}
\usepackage{comment}
\usepackage{subcaption}
\usepackage{float}
\usepackage{hyperref}

\begin{document}
\begin{frontmatter}

\title{Analysis of Structure and Interactions Between Chemical Reactions, Species Transport and Heat Release in Laminar Flames}

\author{Liang Ji \corref{cor1}}
\author{Kalyanasundaram Seshadri \corref{cor1}}
\cortext[cor1]{Corresponding author: kseshadri@ucsd.edu, lji@ucsd.edu}

\address{University of California, San Diego, La Jolla, California 92093-0411, USA}

\date{}

\begin{abstract} 

\newcommand{\eqn}{Eq.}
\newcommand{\eqns}{Eqs.}
\newcommand{\aea}{activation-energy asymptotic}
\newcommand{\caea}{Activation-energy asymptotic}
\newcommand{\caeo}{Activation-energy}
\newcommand{\aeo}{activation-energy}
\newcommand{\aeays}{activation-energy asymptotic analyses}
\newcommand{\rra}{rate-ratio asymptotic}
\newcommand{\crra}{Rate-ratio asymptotic}
\newcommand{\asyan}{asymptotic analysis}
\newcommand{\asyans}{asymptotic analyses} 
\newcommand{\ci}{chemical inhibition}
\newcommand{\chc}{characteristic}
\newcommand{\sr}{strain rate}
\newcommand{\sdr}{scalar dissipation rate}
\newcommand{\fl}{flame}
\newcommand{\cf}{counterflow}
\newcommand{\cp}{c_{\rm {p}}}
\newcommand{\cfg}{counterflowing}
\newcommand{\df}{diffusion flame}
\newcommand{\ppf}{partially-premixed}
\newcommand{\cppf}{Partially-premixed}
\newcommand{\pf}{premixed flame}
\newcommand{\pfs}{premixed flames}
\newcommand{\np}{nonpremixed}
\newcommand{\npf}{nonpremixed flame}
\newcommand{\npfs}{nonpremixed flames}
\newcommand{\pheat}{preheat zone}
\newcommand{\post}{post-flame zone}
\newcommand{\zeld}{Zel'dovich number}
\newcommand{\as}{air stream}
\newcommand{\fs}{fuel stream}
\newcommand{\frs}{fuel-rich stream}
\newcommand{\fls}{fuel-lean stream}
\newcommand{\fsd}{fuel side}
\newcommand{\frbd}{fuel-rich boundary}
\newcommand{\flbd}{fuel-lean boundary}
\newcommand{\fbd}{fuel boundary}
\newcommand{\os}{oxidizer stream}
\newcommand{\osd}{oxidizer side}
\newcommand{\obd}{oxidizer boundary}
\newcommand{\ma}{methane air} 
\newcommand{\ck}{chemical-kinetic}
\newcommand{\cbk}{chain-breaking}
\newcommand{\cbr}{chain-branching}
\newcommand{\gl}{global}
\newcommand{\ggl}{Global}
\newcommand{\me}{mechanism}
\newcommand{\mes}{mechanisms}
\newcommand{\sk}{skeletal}
\newcommand{\rd}{reduced}
\newcommand{\sts}{steady-state}
\newcommand{\nsts}{nonsteady-state}
\newcommand{\pe}{partial equilibrium}
\newcommand{\ff}{flow-field}
\newcommand{\il}{inner layer}
\newcommand{\ol}{oxidation layer}
\newcommand{\dam}{Damk{\"{o}}hler number}
\newcommand{\dams}{Damk{\"{o}}hler numbers}
\newcommand{\fcl}{fuel-consumption layer}
\newcommand{\neql}{nonequilibrium layer}
\newcommand{\rcl}{radical-consumption layer}
\newcommand{\dfl}{diffusion-flame layer}
\newcommand{\rnl}{radical-nonequilibrium layer}
\newcommand{\z}{{\xi}}
\newcommand{\zht}{{\xi}_{\rm {H_2}}}
\newcommand{\aig}{auto-ignition}
\newcommand{\aq}{a_{\rm {e}}}
\newcommand{\zst}{\xi_{\rm {st}}}
\newcommand{\xst}{x_{\rm {st}}}
\newcommand{\xp}{x_{\rm {p}}}
\newcommand{\zhtst}{{\xi}_{\rm {H_2,st}}}
\newcommand{\zp}{\xi_{\rm {p}}}
\newcommand{\zhtp}{{\xi}_{\rm {H_2,p}}}
\newcommand{\pchi}{\chi_{\rm {p}}}
\newcommand{\ipchi}{\chi_{\rm {p}}^{-1}}
\newcommand{\qpchi}{\chi_{\rm {p,q}}}
\newcommand{\stchi}{\chi_{\rm {st}}}
\newcommand{\nchi}{{\chi}^0}
\newcommand{\qstchi}{\chi_{\rm {st,q}}}
\newcommand{\inchi}{\chi_{\rm {I}}}
\newcommand{\zcp}{\xi_{\rm {p}}}
\newcommand{\zcr}{\xi_{\rm {r}}}
\newcommand{\tn}{T^0}
\newcommand{\tu}{T_{\rm {u}}}
\newcommand{\tc}{T_{\rm {c}}}
\newcommand{\yiu}{Y_{\rm {iu}}}
\newcommand{\tb}{T_{\rm {b}}}
\newcommand{\tst}{T_{\rm {st}}}
\newcommand{\tp}{T_{\rm {p}}}
\newcommand{\tfr}{T_{\rm {u}}}
\newcommand{\tfrr}{T_{\rm {ref}}}
\newcommand{\dtref}{{\Delta}T_{\rm {ref}}}
\newcommand{\tr}{T_{\rm {r}}}
\newcommand{\tl}{T_{\rm {l}}}
\newcommand{\mth}{methane}
\newcommand{\heptane}{{\it {n-}}heptane}
\newcommand{\decane}{{\it {n-}}decane}
\newcommand{\isooctane}{{\it {iso-}}octane}
\newcommand{\cisooctane}{{\it {iso-}}Octane}
\newcommand{\nbuta}{{\it {n-}}butanol}
\newcommand{\isobuta}{{\it {iso-}}butanol}
\newcommand{\isoprop}{{\it {iso-}}propanol}
\newcommand{\ethanol}{ethanol}
\newcommand{\ceth}{Ethanol}
\newcommand{\dme}{dimethyl ether}
\newcommand{\cdme}{Dimethyl ether}
\newcommand{\sheptane}{C$_7$H$_{16}$}
\newcommand{\sdecane}{C$_{10}$H$_{22}$}
\newcommand{\soctane}{C$_8$H$_{18}$}
\newcommand{\sdme}{CH$_3$OCH$_3$}
\newcommand{\seth}{C$_2$H$_5$OH}
\newcommand{\sisobuta}{C$_4$H$_8$OH}
\newcommand{\htt}{H$_2$}
\newcommand{\chf}{CH$_4$}
\newcommand{\nt}{N$_2$}
\newcommand{\hto}{H$_2$O}
\newcommand{\ssp}{$\,$}
\newcommand{\smo}{s$^{-1}$}

\newcommand{\tothe}[2]{#1$^{#2}$}
\newcommand{\eti}{\eta_{\rm {i}}}
\newcommand{\cci}{C_{\rm i}}
\newcommand{\echi}{H_{\rm i}}
\newcommand{\cpi}{c_{\rm p,i}}
\newcommand{\cpn}{c^0_{\rm p}}
\newcommand{\cm}{C_{\rm M}}
\newcommand{\cmn}{C_{\rm M}^0}
\newcommand{\cyi}{Y_{\rm {i}}}
\newcommand{\cyin}{Y_{\rm {I}}}
\newcommand{\cyott}{Y_{\rm {O_2,2}}}
\newcommand{\cyotr}{Y_{{\rm {O_2}},r}}
\newcommand{\cyotl}{Y_{{\rm {O_2}},l}}
\newcommand{\cyotst}{Y_{\rm {O_2,st}}}
\newcommand{\cyotln}{Y^0_{{\rm {O_2}},l}}
\newcommand{\cyot}{Y_{\rm {O_2}}}
\newcommand{\cyntot}{Y_{\rm {N_2O,2}}}
\newcommand{\cyf}{Y_{\rm {F}}}
\newcommand{\cyff}{Y_{\rm {F,1}}}
\newcommand{\cyhtf}{Y_{\rm {H_2,1}}}
\newcommand{\cxhtf}{X_{\rm {H_2,1}}}
\newcommand{\cyhtt}{Y_{\rm {H_2,2}}}
\newcommand{\cxhtt}{X_{\rm {H_2,2}}}
\newcommand{\cxi}{X_{\rm {i}}}
\newcommand{\sxi}{x_{\rm {i}}}
\newcommand{\cxf}{X_{\rm {F}}}
\newcommand{\cxfo}{X_{\rm {F,1}}}
\newcommand{\cxft}{X_{\rm {F,2}}}
\newcommand{\sxf}{x_{\rm {F}}}
\newcommand{\sxh}{x_{\rm {H}}}
\newcommand{\sxnto}{x_{\rm {N_2O}}}
\newcommand{\sxnt}{x_{\rm {N_2}}}
\newcommand{\syf}{y_{\rm {F}}}
\newcommand{\syh}{y_{\rm {H}}}
\newcommand{\cxot}{X_{\rm {O_2}}}
\newcommand{\cxnt}{X_{\rm {N_2}}}
\newcommand{\cxntt}{X_{\rm {N_2,2}}}
\newcommand{\cxott}{X_{\rm {O_2,2}}}
\newcommand{\cxotr}{X_{{\rm {O_2}},r}}
\newcommand{\cxotl}{X_{{\rm {O_2}},l}}
\newcommand{\cxotln}{X^0_{{\rm {O_2}},l}}
\newcommand{\cxntot}{X_{\rm {N_2O,2}}}
\newcommand{\cxntoo}{X_{\rm {N_2,1}}}
\newcommand{\cxnto}{X_{\rm {N_2O}}}
\newcommand{\sxot}{x_{\rm {O_2}}}
\newcommand{\echot}{H_{\rm {O_2}}}
\newcommand{\echnto}{H_{\rm {N_2O}}}
\newcommand{\echnt}{H_{\rm {N_2}}}
\newcommand{\echhto}{H_{\rm {H_2O}}}
\newcommand{\echcot}{H_{\rm {CO_2}}}
\newcommand{\echco}{H_{\rm {CO}}}
\newcommand{\echf}{H_{\rm {F}}}
\newcommand{\cxcht}{X_{\rm {CH_3}}}
\newcommand{\cxh}{X_{\rm {H}}}
\newcommand{\cxo}{X_{\rm {O}}}
\newcommand{\cxoh}{X_{\rm {OH}}}
\newcommand{\cxht}{X_{\rm {H_2}}}
\newcommand{\sxht}{x_{\rm {H_2}}}
\newcommand{\sxhtn}{(x_{\rm {H_2}})^0}
\newcommand{\cxhto}{X_{\rm {H_2O}}}
\newcommand{\sxhto}{x_{\rm {H_2O}}}
\newcommand{\cxhtop}{X_{\rm {H_2O,p}}}
\newcommand{\cxhtost}{X_{\rm {H_2O,st}}}
\newcommand{\cxcot}{X_{\rm {CO_2}}}
\newcommand{\sxcot}{x_{\rm {CO_2}}}
\newcommand{\cxcotp}{X_{\rm {CO_2,p}}}
\newcommand{\cxcotst}{X_{\rm {CO_2,st}}}
\newcommand{\cxcop}{X_{\rm {CO,p}}}
\newcommand{\cxntop}{X_{\rm {N_2O,p}}}
\newcommand{\cxco}{X_{\rm {CO}}}
\newcommand{\sxco}{x_{\rm {CO}}}
\newcommand{\cxotp}{X_{\rm {O_2,p}}}
\newcommand{\cyotp}{Y_{\rm {O_2,p}}}
\newcommand{\cxhtn}{X_{\rm {H_2}}^0}
\newcommand{\cxcon}{X_{\rm {CO}}^0}
\newcommand{\cxnton}{X_{\rm {N_2O}}^0}
\newcommand{\syi}{y_{\rm {i}}}
\newcommand{\sycht}{y_{\rm {CH_3}}}
\newcommand{\syht}{y_{\rm {H_2}}}
\newcommand{\syot}{y_{\rm {O_2}}}
\newcommand{\sznto}{{z}_{\rm {N_2O}}}
\newcommand{\szco}{{z}_{\rm {CO}}}
\newcommand{\szot}{{z}_{\rm {O_2}}}
\newcommand{\szn}{{z}^0}
\newcommand{\szcon}{{z}_{\rm {CO}}^0}
\newcommand{\szotn}{{z}_{\rm {O_2}}^0}
\newcommand{\szcoc}{{z}_{\rm {CO,c}}}
\newcommand{\szotc}{{z}_{\rm {O_2,c}}}
\newcommand{\sznton}{{z}_{\rm {N_2O}}^0}
\newcommand{\szht}{{z}_{\rm {H_2}}}
\newcommand{\szhtn}{{z}_{\rm {H_2}}^0}
\newcommand{\szhtc}{{z}_{\rm {H_2},c}}
\newcommand{\stn}{t^0}
\newcommand{\sztht}{\tilde{z}_{\rm {H_2}}}
\newcommand{\sztco}{\tilde{z}_{\rm {CO}}}
\newcommand{\sztcon}{\tilde{z}_{\rm {CO}}^0}
\newcommand{\tzeta}{\tilde{\zeta}}
\newcommand{\cdi}{D_{\rm {i}}}
\newcommand{\lei}{Le_{\rm {i}}}
\newcommand{\lef}{Le_{\rm {F}}}
\newcommand{\lecht}{Le_{\rm {CH_3}}}
\newcommand{\leot}{Le_{\rm {O_2}}}
\newcommand{\lento}{Le_{\rm {N_2O}}}
\newcommand{\leht}{Le_{\rm {H_2}}}
\newcommand{\leco}{Le_{\rm {CO}}}
\newcommand{\leh}{Le_{\rm {H}}}
\newcommand{\lehto}{Le_{\rm {H_2O}}}
\newcommand{\lecot}{Le_{\rm {CO_2}}}
\newcommand{\wbar}{\widehat{W}}
\newcommand{\wi}{W_{\rm {i}}}
\newcommand{\wf}{W_{\rm {F}}}
\newcommand{\win}{W_{\rm {I}}}
\newcommand{\wot}{W_{\rm {O_2}}}
\newcommand{\wnt}{W_{\rm {N_2}}}
\newcommand{\wnto}{W_{\rm {N_2O}}}
\newcommand{\wn}{w_{\rm {n}}}
\newcommand{\wk}{w_{\rm {k}}}
\newcommand{\omk}{{\omega}_{\rm {k}}}
\newcommand{\om}{\omega}
\newcommand{\kn}{k_{\rm {n}}}
\newcommand{\ckn}{K_{\rm {n}}}
\newcommand{\cqk}{Q_{\rm {k}}}
\newcommand{\nik}{\nu_{\rm {ik}}}
\newcommand{\nrho}{{\rho}^0}
\newcommand{\ptau}{\tau_{\rm {p}}}
\newcommand{\sttau}{\tau_{\rm {st}}}
\newcommand{\ntau}{\tau^0}
\newcommand{\p}{\rm {p}}
\newcommand{\qh}{q_{\rm {H}}}
\newcommand{\qf}{q_{\rm {F}}}
\newcommand{\qht}{q_{\rm {H_2}}}
\newcommand{\qco}{q_{\rm {CO}}}
\newcommand{\qnto}{q_{\rm {N_2O}}}
\newcommand{\sznls}{{\hat{z}}^0}
\newcommand{\inls}{\psi}
\newcommand{\szhtls}{{\hat{z}}_{\rm {H_2}}}
\newcommand{\szhtnls}{{\hat{z}}_{\rm {H_2}}^0}
\newcommand{\szcols}{{\hat{z}}_{\rm {CO}}}
\newcommand{\cxhtnls}{{\hat{X}}_{\rm {H_2}}^0}
\newcommand{\cxhtnss}{{\bar{X}}_{\rm {H_2}}^0}
\newcommand{\sznss}{\tilde{z}^0}
\newcommand{\szhtnss}{{\bar{z}}_{\rm {H_2}}^0}
\newcommand{\etac}{{\eta}_{\rm {c}}}
\newcommand{\neta}{{\eta}^0}
\newcommand{\aaa}{A}
\newcommand{\rrrt}{R}
\newcommand{\rrr}{R^0}
\newcommand{\eigenvalue}{\omega}
\newcommand{\cshat}{\overline{S}}
\newcommand{\cyhep}{Y_{\rm {hep}}}
\newcommand{\cyoct}{Y_{\rm {oct}}}
\newcommand{\cydme}{Y_{\rm {dme}}}
\newcommand{\cyeth}{Y_{\rm {eth}}}
\newcommand{\cyhepo}{Y_{\rm {hep,1}}}
\newcommand{\cyocto}{Y_{\rm {oct,1}}}
\newcommand{\cydmeo}{Y_{\rm {dme,1}}}
\newcommand{\cydmer}{Y_{{\rm {dme}},r}}
\newcommand{\cydmel}{Y_{{\rm {dme}},l}}
\newcommand{\cydmern}{Y^0_{{\rm {dme}},r}}
\newcommand{\cyetho}{Y_{\rm {eth,1}}}
\newcommand{\rrhep}{w_{\rm {hep}}}
\newcommand{\rroct}{w_{\rm {oct}}}
\newcommand{\rrdme}{w_{\rm {dme}}}
\newcommand{\rreth}{w_{\rm {eth}}}
\newcommand{\molhep}{W_{\rm {hep}}}
\newcommand{\moloct}{W_{\rm {oct}}}
\newcommand{\moldme}{W_{\rm {dme}}}
\newcommand{\moleth}{W_{\rm {eth}}}
\newcommand{\molot}{W_{\rm {O_2}}}
\newcommand{\diffhep}{D_{\rm {hep}}}
\newcommand{\diffoct}{D_{\rm {oct}}}
\newcommand{\diffeth}{D_{\rm {eth}}}
\newcommand{\diffdme}{D_{\rm {dme}}}
\newcommand{\diffot}{D_{\rm {O_2}}}
\newcommand{\cqhep}{Q_{\rm {hep}}}
\newcommand{\cqoct}{Q_{\rm {oct}}}
\newcommand{\cqdme}{Q_{\rm {dme}}}
\newcommand{\cqeth}{Q_{\rm {eth}}}
\newcommand{\zhep}{{\xi}_{\rm {hep}}}
\newcommand{\zoct}{{\xi}_{\rm {oct}}}
\newcommand{\zdme}{{\xi}_{\rm {dme}}}
\newcommand{\zeth}{{\xi}_{\rm {eth}}}
\newcommand{\zhepst}{{\xi}_{\rm {hep,st}}}
\newcommand{\zoctst}{{\xi}_{\rm {oct,st}}}
\newcommand{\zdmest}{{\xi}_{\rm {dme,st}}}
\newcommand{\zethst}{{\xi}_{\rm {eth,st}}}
\newcommand{\lehep}{Le_{\rm {hep}}}
\newcommand{\leoct}{Le_{\rm {oct}}}
\newcommand{\ledme}{Le_{\rm {dme}}}
\newcommand{\leeth}{Le_{\rm {eth}}}
\newcommand{\cxhep}{X_{\rm {hep}}}
\newcommand{\cxoct}{X_{\rm {oct}}}
\newcommand{\cxdme}{X_{\rm {dme}}}
\newcommand{\cxdmer}{X_{{\rm {dme}},r}}
\newcommand{\cxdmel}{X_{{\rm {dme}},l}}
\newcommand{\cxdmern}{X^0_{{\rm {dme}},r}}
\newcommand{\cxeth}{X_{\rm {eth}}}
\newcommand{\cxhepo}{X_{\rm {hep,1}}}
\newcommand{\cxocto}{X_{\rm {oct,1}}}
\newcommand{\cxdmeo}{X_{\rm {dme,1}}}
\newcommand{\cxetho}{X_{\rm {ethn,1}}}
\newcommand{\lphi}{{\phi}_l}
\newcommand{\rphi}{{\phi}_r}
\newcommand{\ssr}{a}
\newcommand{\qssr}{{\ssr}_{\mathrm {q}}}
\newcommand{\qssrpp}{{\ssr}_{\mathrm {q,pp}}}
\newcommand{\qssrnp}{{\ssr}_{\mathrm {q,np}}}
\newcommand{\sdistance}{l}
\newcommand{\cvelr}{V_{\mathrm {r}}}
\newcommand{\cvell}{V_{\mathrm {l}}}
\newcommand{\cvelf}{V_{\mathrm {1}}}
\newcommand{\cvelo}{V_{\mathrm {2}}}
\newcommand{\denr}{{\rho}_{\mathrm {r}}}
\newcommand{\denl}{{\rho}_{\mathrm {l}}}
\newcommand{\denf}{{\rho}_{\mathrm {1}}}
\newcommand{\deno}{{\rho}_{\mathrm {2}}}
\newcommand{\cyfr}{Y_{\mathrm {F,r}}}
\newcommand{\cyfl}{Y_{\mathrm {F,l}}}
\newcommand{\ffd}{flow-field}
\newcommand{\ppm}{partially premixed}
\newcommand{\cppm}{Partially premixed}
\newcommand{\od}{oxidizer-duct}
\newcommand{\ob}{oxidizer-boundary}
\newcommand{\tempf}{T_{1}}
\newcommand{\tempo}{T_{2}}
\newcommand{\cvlg}{V_{\rm {s}}}
\newcommand{\srf}{{a}_{1}}
\newcommand{\sro}{{a}_{2}}
\newcommand{\tg}{T_{\rm {s}}}
\newcommand{\jfs}{j_{\rm F,s}}
\newcommand{\jis}{j_{i,s}}
\newcommand{\cyis}{Y_{i,s}}
\newcommand{\brate}{{\dot{m}}}
\newcommand{\cyfg}{Y_{\rm F,s}}
\newcommand{\cxfg}{X_{\rm F,s}}
\newcommand{\tig}{T_{\rm {ig}}}

\newcommand{\hyt}{\ce{H2}}
\newcommand{\co}{\ce{CO}}
\newcommand{\ot}{\ce{O2}}
\newcommand{\hco}{\ce{HCO}}
\newcommand{\oh}{\ce{OH}}
\newcommand{\hot}{\ce{HO2}}
\newcommand{\htot}{\ce{H2O2}}
\newcommand{\chthr}{\ce{CH3}}
\newcommand{\cthtr}{\ce{C2H3}}
\newcommand{\cthfo}{\ce{C2H4}}
\newcommand{\cthf}{\ce{C2H5}}
\newcommand{\chto}{\ce{CH2O}}
\newcommand{\cthforooh}{\ce{C2H4OOH}}
\newcommand{\ctrhsvn}{\ce{NC3H7}}
\newcommand{\ctrhs}{\ce{C3H6}}
\newcommand{\ncsvnoqooh}{\ce{NC7OQOOH}}
\newcommand{\ethal}{\ce{C2H5OH}}
\newcommand{\chthrchoh}{\ce{CH3CHOH}}
\newcommand{\chtchtoh}{\ce{CH2CH2OH}}
\newcommand{\chthrchto}{\ce{CH3CH2O}}
\newcommand{\chthrcho}{\ce{CH3CHO}}


A novel method for analyzing counterflow diffusion flames, inspired by Zurada's sensitivity approach for neural networks, is proposed to identify critical species influencing the heat release rate in combustion. By further analyzing concentration changes of selected key species and radicals, this method reveals complex interactions among them across regions with temperature. To illustrate this approach, the study investigates the mechanisms of auto-ignition of n-heptane and ethanol mixtures in a counterflow configuration under low strain rates. In mixtures where n-heptane is dominant, the inhibition of low-temperature chemistry (LTC) by addition of  ethanol impacts the heat release rate in regions where the temperature is higher through the diffusion of specific radicals such as \chto, \cthfo, \ctrhs, and \htot. In mixtures where ethanol is dominant, the high ethanol fractions in the mixture increase the heat release rate, primarily due to ethanol decomposition and its subsequent reactions. This method effectively quantifies and compares the influence of both chemical kinetics and species diffusion effects, providing detailed insights into the interactions among species across the reactive field when analyzing the counterflow configuration of complex fuel mixtures.
\end{abstract}

\end{frontmatter}

\clearpage

\section {Introduction}\addvspace{10pt}
The laminar, steady, one-dimensional  counterflow diffusion flame is investigated to elucidate interaction between diffusion and chemistry\cite{TSUJI198293}. 
Through the counterflow configuration, Seshadri \cite{seshadri:1978:laminar} studied extinction of diffusion flame methanol, heptane and wood in the presence of suppressive agents such as nitrogen and water. Seiser et\ al. \cite{SEISER20002029} elucidated the mechanisms of extinction and autoignition of n-heptane, finding that strain has greater influence on low-temperature chemistry than the temperature of the reactants. Ji et\ al. \cite{JI20232007} investigated the impact of iso-butanol and ethanol on the auto-ignition of n-decane and n-heptane. It is observed that addition of iso-butanol and ethanol to n-decane or n-heptane elevated the auto-ignition temperature at low strain rates, indicating that iso-butanol and ethanol inhibits the low-temperature chemistry (LTC) of n-decane and n-heptane. These observations were further supported by sensitivity analysis.

Sensitivity analysis have been extensively employed to elucidate the influence of selected parameters on combustion. They are useful for identification and quantification of the role of selected parameters, reveal their predominant controlling influence alongside their indirect effect on changes in their values \cite{Rabitz1983}. They play a crucial role in uncertainty analysis, estimation of parameter, and investigation or reduction of mechanism \cite{TURANYI199741}. This approach is instrumental in understanding the sensitivity of the predicted outcomes or quantities of interest (QoIs) to uncertain parameters \cite{Kalen2015}. For example, sensitivity analysis facilitates quantification of the indirect influence of the rate constants of reactions in terms of temperature. However, the traditional sensitivity analysis in combustion research focuses on the systematic impact of parameters on output variables. It falls short of detailed explaining the direct influence between reactions and the output variables, nor the subsequent effects of these output variables changes on other output variables.

Another widely used method, path flux analysis (PFA) method, plays a crucial role in dissecting the production and consumption fluxes (pathways) in chemical mechanism. It is employed alongside in direct relation graph (DRG) method to identify critical species and reactions. Sun et\ al. demonstrated that the skeletal mechanisms refined by PFA exhibits enhanced accuracy compared to those derived by DRG method with the similar size in several cases\cite{SUN20101298} \cite{Wei2016}. Path flux analysis could explicitly offer insight into the consumption and production of species in zero-dimensional models or single point in one-dimensional flame configuration, facilitating analysis of kinetic mechanism. However, for the counterflow flame configuration, it has difficulty in elucidating the interaction cross different spatial points.

Recently, there have been numerous studies applying deep learning to chemical ordinary differential equations (ODEs). Ji et al.\cite{ji2021stiff} developed the stiff-PINN approach that utilizes QSSA to enable the PINN to solve stiff chemical kinetics. The multiscale physics-informed neural network (MPINN) approach proposed by Weng et al.\cite{weng2022mpinn} is based on the regular physics-informed neural network (PINN) for solving stiff chemical kinetic problems with governing equations of stiff ODEs. Su et al.\cite{SU2023112732} employed a neural ordinary differential equation (Neural ODE) framework to optimize the kinetic parameters of reaction mechanisms, showing that the proposed algorithm can optimize stiff chemical models with sufficient accuracy, efficiency, and robustness. The forward propagation in neural networks shows a kind of equivalence to chemical ODEs. Thus, it is also possible to introduce sensitivity analysis methods for neural networks to chemical systems.

Zurada's sensitivity method, widely used in the analysis of neural networks for the reduction of training set size\cite{ZURADA1997177}, employs calculations based on partial derivatives with respect to input variables for redundant feature selection or deletion. Inspired by Zurada's method and considering the unique characteristics of combustion chemistry alongside the governing equations of the counterflow configuration, we introduce a supplemental method. This method, grounded in the use of partial derivatives, aims to analyze the interplay across different temperature zones in counterflow flames. It is employed to provide a detailed explanation of the interactions between n-heptane and ethanol in their binary mixtures in counterflow diffusion flames.

\section {Methodology}
\subsection {Analysis of reaction rate change }\addvspace{10pt}
The objective of this analysis is to identify factors, either changes in value of the rate constant, species concentration or both, that primarily contribute to variations of the reaction rate under selected conditions.

The rates of forward reaction $\dot {\omega}_{f,k}$ and reverse reaction $\dot {\omega}_{b,k}$ of the $k^{th}$ reversible reaction are determined by the product of the concentration of species $i$, $c_i$  and their respective rate constants $k_{f,k}$ and $k_{b,k}$. Hence, $\dot{\omega}_{f,k} = k_{f,k} \prod_{i=1}^{m} c_i^{\nu'_i}$, $\dot{\omega}_{b,k} = k_{b,k} \prod_{i=1}^{m} c_i^{\nu''_i}$, 
where the parameters $\nu'_i$, $\nu''_i$ represent the stoichiometric coefficients for species ${i}$ appearing as a reactant and as a product in a reversible reaction, respectively. 

In the analysis described here, forward and reverse steps of an elementary reaction is considered to be two separate reactions. Thus if there are $M$ reversible reactions, the total number of reactions considered as $2 M$. \\
\begin{equation}
  \dot{\omega}_n  = 
    \begin{cases} 
    \displaystyle \; \dot{\omega}_{f,k}, n=2k-1,\text{(for forward step)}\\
    \displaystyle \dot{\omega}_{b,k}, n=2k,\text{(for reverse step)}
    \end{cases} \text{where}{\hspace{2mm}} \textit{k =1,2 ...M}
\label{eq:sep}
\end{equation}
Here, \textit{k} denotes the $k^{th}$ reversible elementary reactions before their separation into forward and reverse steps, while \textit{n} represents the $n^{th}$ reactions after separation. The total number of reactions after separation is $N$.

Similar to the approach taken in sensitivity analysis, the calculation of partial derivatives of the $n^{th}$ reaction rate, $\dot {\omega}_n$, with respect to the rate constant, $k_n$, and species concentration, $c_i$, is a key parameter in reaction-rate-change-analysis. 

Let $\Delta \dot {\omega}_n$ represent the difference in the value of $\dot {\omega}_n$ as a result of changes in the input variables, for example changes in initial composition of the reactive mixture. Let the corresponding changes in value of the rate constant $k_n$ and concentration of species $i$ be $\Delta k_n$ and $\Delta c_i$, respectively. It follows that,
\begin {equation}
\begin{array}{l}
\displaystyle \Delta \dot {\omega}_n| _{k_n} =  \frac{\Delta \dot {\omega}_n}{\Delta \dot{\omega}^{\text{approx}}_n} \times {\displaystyle \frac {\partial \dot {\omega}_n}{\partial k_n} \times \Delta k_n}   \vspace{4mm} \\
\displaystyle \Delta \dot {\omega}_n| _{c_i} = \frac{\Delta \dot {\omega}_n}{\Delta \dot{\omega}^{\text{approx}}_n} \times {\displaystyle \frac{\partial \dot{\omega}_n}{\partial c_i} \times \Delta c_i} 
\end{array}
\end{equation}
where $\Delta \dot {\omega}_n| _{k_n}$ and $ \Delta \dot {\omega}_n| _{c_i}$ are change of reaction rate caused by $\Delta k_n$ and $\Delta c_i$. And the derivatives ${\Delta \dot{\omega}^{\text{approx}}_n}$, ${\partial \dot {\omega}_n}/{\partial k_n}$ and ${\partial \dot {\omega}_n}/{\partial c_i}$ are given by:
\begin {equation}
\begin{array}{l}
{\Delta \dot{\omega}^{\text{approx}}_n} = {\displaystyle \frac{\partial \dot {\omega}_n}{\partial k_n} \times \Delta k_n + \sum_{i=1}^{m} \left (\frac{\partial \dot {\omega}_n}{\partial c_i } \times \Delta c_i \right)}  \vspace{4mm} \\
{\displaystyle \frac{\partial \dot {\omega}_n}{\partial k_n} = \prod_{i=1}^{m} c_i^{\lvert \nu_i \rvert}, \frac{\partial \dot {\omega}_n}{\partial c_i} = k_n \nu_i  c_i^{\lvert \nu_i \rvert- 1} \prod_{\substack{j=1 \\ i \neq j}}^{m} {c_j}^{\lvert \nu_j \rvert}}
\end{array}
\end{equation}
with $\nu_i = \nu''_i - \nu'_i$. Here, $m$ represents the total number of species involved in the $n^{th}$ reaction. The term ${\Delta \dot{\omega}^{\text{approx}}_n}$ is an approximation of ${\Delta \dot{\omega}_n}$, derived from the Taylor expansion truncated at the first-order derivative. A second-order Taylor series expansion can be employed if improved accuracy is desired in evaluation of ${\Delta \dot {\omega}_n| _{c_i}}$ and ${\Delta \dot {\omega}_n| _{k_n}}$. Details are provided in \ref{Appendix:b1}.
\subsection {Analysis of heat release rate change }\addvspace{10pt}
The goal of the heat-release-rate-analysis is to provide quantitative information concerning the impact of changes in concentration each species on overall heat release rate, $\dot{Q}$, under different input boundary conditions. The quantity $\dot{Q}=\sum_k{\Delta \text{H}_k \times \dot {\omega}_k}$, where $\Delta \text{H}_k$ and $\dot {\omega}_k$ are, respectively, the enthalpy change and net reaction rate of the $k^{th}$ reaction. Furthermore, the net reaction rate $\dot {\omega}_k = \dot {\omega}_{f,k}- \dot {\omega}_{b,k}$.  Hence, through separation operation, Eqn.\ \eqref{eq:sep}, it follows that
\begin{equation}
  \dot{Q} =   -\sum_n{\Delta \text{H}_n \times \dot {\omega}_n}
\end{equation}

Heat-release-rate-analysis presented here is primarily focused on the high-temperature region where the input enthalpy of the reactant streams is maintained at constant values. Therefore, the local temperature at selected points for different boundary conditions of fuel streams are nearly the same allowing for the reasonable assumption that reaction enthalpy $\Delta \text{H}_n$, which depends on local temperature, remains constant. Therefore, the contribution to change of heat release rate, $\Delta \dot {Q}$, arising from  changes in rate constant, $\Delta k_n$ and species concentrations $\Delta c_i$, are written as follows:
 
\begin{equation}
\displaystyle \Delta \dot {Q}| _{k,n} =  -\Delta \text{H}_n\times \Delta \dot {\omega}_n|_{k,n}
\end{equation}

\begin{equation}
\displaystyle \Delta \dot {Q}| _{c_i} = \displaystyle \sum_n \left( -\Delta \text{H}_n \times \Delta \dot {\omega}_n|_{c_i}\right)
\end{equation}

The terms ${\Delta \dot {Q}}| _{k,n}$,  and ${\Delta \dot {Q}}|_{c_i}$ represent the change of heat release rate attributed to the change in value of rate constant $\Delta k_n$ of the $n^{th}$ reaction and the change in value of concentration of species $i$, $\Delta c_i$.
\subsection{Analysis of species concentration change}\addvspace{10pt}
The goal of the analysis of species concentration change is to provide quantitative information  concerning the impact of changes in rates of reactions on changes in concentration of species and elucidate the interaction among various species, particularly for those for which steady-state approximation are reasonably accurate. For species that do not satisfy steady-state approximation, the analysis includes the effects of species diffusion and convection on concentration.  Furthermore, diagrams illustrating species governing equation terms provides a visual representation that enhances understanding interactions associated with these species.

As an example for the counterflow flame the steady state balance equation for species $i$ is given by :
\begin{equation}
\displaystyle
0=-{\rho}u\frac{dY_i}{dz}-\frac{d {j_i}}{dz}+{W_i  (\dot{\omega}_i^+ -  \dot{\omega}_i^-})
\label{eq:gvn_eq}
\end{equation}
Here, $z$ represents the spatial co-ordinate, $\rho$ the density, $u$ the mass-averaged velocity, $Y_i$, and  $W_i$ are the mass fraction and molecular weight of species $i$, $j_i$  the diffusive flux of species $i$, and $\dot{\omega}_i^+$ and  $\dot{\omega}_i^-$ are, respectively the rate of production and rate of consumption of species $i$.  The first term on the right side of Eqn.\ \eqref{eq:gvn_eq} represents convective transport, the second term diffusive transport, and the third term the net rate of production of species $i$.
From Eqn.\ \eqref{eq:gvn_eq} and expression obtained for $c_i$ is obtained and written as
\begin{equation}
 c_i =\frac{\displaystyle -{{\rho}u\frac{dY_i}{dz}}/{W_i} -{\frac{d {j_i}}{dz}}/{W_i}+ \sum_{k} \lvert \nu_{i,k} \rvert \left[(1-\delta'_{i,k})  \dot{\omega}_{f,k} + \delta'_{i,k}  \dot{\omega}_{b,k} \right]}  {\displaystyle\sum_{k} \lvert\nu_{i,k}\rvert \left[\delta'_{i,k} w^{den}_{f,k}+(1-\delta'_{i,k})w^{den}_{b,k}\right]}
\label{eq:xb2}
\end{equation}
Here, $\delta'_{i,k}$ defined as
\begin{equation}
\delta'_{i,k}  =
\begin{cases}
1 & \text{if } \nu_{i,k}  <0, \\
0 &  \text{if } \nu_{i,k} > 0
\end{cases}
\end{equation}
The term $\nu_{i,k}$ is given by $\nu''_{i,k} - \nu'_{i,k}$, where $\nu'_{i,k}$, $\nu''_{i,k}$ are the stoichiometric coefficients for species $i$ appearing as a reactant and as a product in $k^{th}$ reaction, respectively. The detailed derivation of Eqn.\ \eqref{eq:xb2} is shown in \ref{Appendix:b2}. 
In Eqn.\ \eqref{eq:xb2}, the terms $[-{\rho}u\frac{dY_i}{dz}/W_i]$ and $[-\frac{d {j_i}}{dz}/W_i]$ correspond to convective and diffusive mass transfer related term of species $i$, respectively. These terms can be neglected if steady-state approximation is satisfied. The term ${\sum_{k} \lvert \nu_{i,k} \rvert \left[(1-\delta'_{i,k})  \dot{\omega}_{f,k} + \delta'_{i,k}  \dot{\omega}_{b,k} \right]}$ accounts for all reactions producing species $i$ while ${\sum_{k} \lvert\nu_{i,k}\rvert \left[\delta'_{i,k} w^{den}_{f,k}+(1-\delta'_{i,k})w^{den}_{b,k}\right]}$ accounts for all reactions consumed species $i$.\\ 

The framework described by Eqn.\ \eqref{eq:xb2}, is used to  analyze changes in the concentration of species $i$. For the $k^{th}$ elemental reaction,  the contribution of its forward reaction rate with respect to species $ c_i$ is:
\begin{equation}
 \Delta c_i|_{f, k} =  
   \begin {cases} \displaystyle
   \displaystyle \frac{1}{\text{sum}} \times \frac{ \partial  c_i}{\partial  w_{f,k}} \times \Delta w_{f,k}  {\,},  \nu_{i,k}  >0\\
   \displaystyle \frac{1}{\text{sum}} \times \frac{ \partial  c_i}{\partial w^{den}_{f,k}} \times \Delta w^{den}_{f,k}, \nu_{i,k}  <0
   \end{cases}
\end{equation}
Similarly, its backward reaction rate contribution with respect to $ c_i$ is expressed as:
\begin{equation}
  \Delta c_i|_{b, k} = 
    \begin{cases} \displaystyle
    \displaystyle \frac{1}{\text{sum}} \times \frac{\partial  c_i}{\partial w_{b,k} } \times \Delta w_{b,k} {\,}, \nu_{i,k}  <0 \\
    \displaystyle \frac{1}{\text{sum}} \times \frac{\partial  c_i}{\partial w^{den}_{b,k}} \times \Delta w^{den}_{b,k}, \nu_{i,k}  >0
    \end{cases}
\end{equation}
The contributions of convective mass transfer with respect to changes in $ c_i$ are identified as:
\begin{equation}
     \Delta c_i|_{\text{convection}} = \displaystyle \frac{1}{\text{sum}} \times \frac{\partial  c_i}{\partial ({\rho}u\frac{dY_i}{dz}) } \times \Delta ({\rho}u\frac{dY_i}{dz})
\end{equation}    
Similarly, the contributions of diffusion mass transfer with respect to changes in $ c_i$ is expressed as:
\begin{equation}
     \Delta c_i|_{\text{diffusion}} = \displaystyle \frac{1}{\text{sum}} \times \frac{ \partial  c_i}{\partial (\frac{d {j_i}}{dz})} \times \Delta (\frac{d {j_i}}{dz})
\end{equation}
Here, $\text{sum}$ defined as the aggregate of all contributions to changes in $ c_i$, encapsulating those from both the forward and backward reaction rates, as well as from the convective and diffusive mass transfers:
\begin{align}
\text{sum} &= \displaystyle \sum \frac{ \partial  c_i}{\partial w_{f,k} } \times \Delta w_{f,k} + \displaystyle \sum \frac{ \partial  c_i}{\partial w^{den}_{f,k}} \times \Delta w^{den}_{f,k} \nonumber \\
&+ \displaystyle \sum \frac{ \partial  c_i}{\partial w^{den}_{f,k}} \times \Delta w^{den}_{f,k} + \displaystyle \sum \frac{\partial  c_i}{\partial w^{den}_{b,k}} \times \Delta w^{den}_{b,k} \nonumber\\
&+ \displaystyle \frac{\partial  c_i}{\partial ({\rho}u\frac{dY_i}{dz}) } \times \Delta ({\rho}u\frac{dY_i}{dz}) + \frac{ \partial  c_i}{\partial (\frac{d {j_i}}{dz})} \times \Delta (\frac{d {j_i}}{dz}) 
\end{align}
It has been established that for a species for which steady-state approximation is accurate, the magnitudes of both the rate of production and the rate of consumption are much larger than the sum of the magnitudes of the convective and diffusive terms \cite{FAW_book}. 
Consequently, upon neglecting the diffusion and convection terms, the concentration of species $i$, as depicted in Eqn.\ \eqref{eq:xb2}, simplifies to:
\begin{equation}
 c_i = \frac{\displaystyle \sum_{k} \lvert \nu_{i,k}\rvert \left[(1-\delta'_{i,k}) w_{f,k} + \delta'_{i,k}w_{b,k}\right]}{\displaystyle \sum_{k} \lvert \nu_{i,k} \rvert \left[\delta'_{i,k}  w^{den}_{f,k}+(1-\delta'_{i,k})w^{den}_{b,k}\right]}
\end{equation}
Furthermore, the term $\text{sum}$ simplifies in the steady-state(ss) approximation as follows:
\begin{align}
\text{sum}_{ss} &=  \displaystyle \sum \frac{ \partial  c_i}{\partial w_{f,k} } \times \Delta w_{f,k} +\displaystyle \sum \frac{ \partial  c_i}{\partial w^{den}_{f,k}} \times \Delta w^{den}_{f,k} \nonumber \\
&+ \displaystyle \sum \frac{ \partial  c_i}{\partial w^{den}_{f,k}} \times \Delta w^{den}_{f,k} + \displaystyle \sum \frac{\partial  c_i}{\partial w^{den}_{b,k}} \times \Delta w^{den}_{b,k} 
\end{align}

In scenarios where steady-state approximation is not applicable to species $c_i$, it becomes critical to consider all terms in the governing equation, including those related to diffusion and convection. This comprehensive approach elucidates the relative significance of diffusion and convection and the chemical reaction term.
Furthermore, for species that do not maintain steady-state, visualizing convection, diffusion and net production terms in Eqn.\ \eqref{eq:gvn_eq} helps to describe scenarios wherein species are produced in one region of the reaction zone transported to a different region of the reaction zone where they react.  This is one of the key differences between reactions in flow systems (for example strained premixed and non-premixed flames) and non-flow systems (for example reactions in shock-tubes). \\

At a specific location, the dominance of one term over others in species equation highlights its pivotal role in shaping concentration profile. A positive value indicates an  enhancement in concentration, whereas a negative value signifies a diminishing effect on concentration profile. This approach is elaborated in results and discussion section.
\section {Demonstration of methodology}  \addvspace{10pt}
Low-temperature chemistry (LTC) is an intrinsic feature of combustion of hydrocarbons such as n-alkanes, alkenes and cycloalkanes, \cite{LIU2004320}. The LTC of n-heptane, for example, has been extensively investigated in various experimental setups, including the counterflow flame \cite{SEISER20002029}, shock tube\cite{HERZLER20051147}, jet-stirred reactor\cite{XIE2022112177}, microgravity droplet flame \cite{FAROUK2014565}. Further studies have explored the impact of alcohol addition to n-heptane or other hydrocarbons in combustion characteristics\cite{ZHANG201331}-\cite{GUO2022}. While, traditional sensitivity method enhances the understanding of interactions within mechanisms, it falls short of detailing the influence between reactions and species across the reactive field. 

Here, we demonstrate the use of the analysis method developed above to elucidate interaction between ethanol and n-heptane in counterflow flame. Computational results from a previous investigation \cite{ji2024experimental} are used to illustrate the method.  This previous study reports on species distribution, flame structure and critical conditions of auto-ignition obtained employing the liquid-fuel counterflow configuration.  These computations are performed using Cantera \cite{cantera:2023} C++ interface with modified boundary conditions at the liquid-gas interface. The mix-average transport model is applied to obtained steady-state solutions. Kinetic modeling is carried out using the San Diego Mechanism \cite{newsandiegomech}.

The fuels tested include n-heptane, ethanol, and their mixtures with volumetric composition of \mbox{20\% n-heptane} + \mbox{80\% ethanol},  \mbox{50\% n-heptane} + \mbox{50\% ethanol},   \mbox{70\% n-heptane} + \mbox{30\% ethanol}, \mbox{80\% n-heptane} + \mbox{20\% ethanol}, and \mbox{90\% n-heptane} + \mbox{10\% ethanol}. The oxidizer is air.

\subsection{Auto-ignition temperature and heat release rate analysis}
\begin {figure} [h!]
\centerline{
 \includegraphics[width = 192pt]{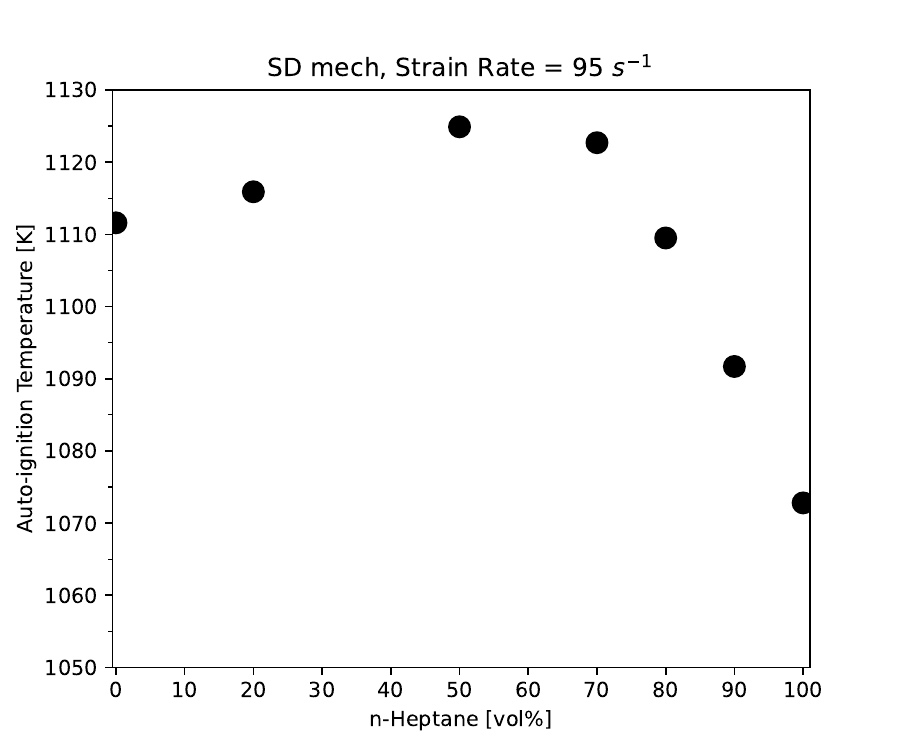}
}
\caption{Auto-ignition temperature at strain rate of 95 s$^{-1}$.}
\label{fig:AT}
\end{figure}

\begin {figure} [h!]
\centerline{\includegraphics[width = 192pt]{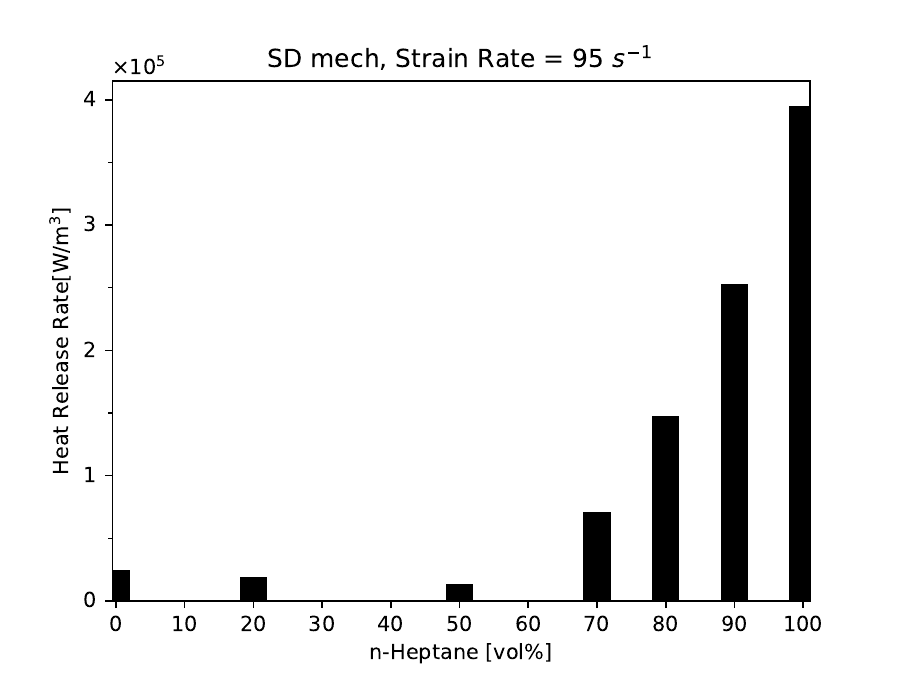}}
\caption{Heat release rate in the high-temperature zone [T$_{\text{ox}}$=1060 K]}
\label{fig:heatrelease}
\end{figure}

Fig.\ \ref{fig:AT} shows auto-ignition temperature calculated at low strain rate, 95$s^{-1}$ for various volume fractions of ethanol in n-heptane, and  Fig.\ \ref{fig:heatrelease} shows the variations in heat release rate, prior to auto-ignition, for these mixtures within the high-temperature zone, particularly focusing at 4.3{\,}mm above the liquid-gas interface and at an oxidizer temperature, T$_\text{ox}$, of 1060K. A comparison between Fig.\ \ref{fig:AT} and Fig.\ \ref{fig:heatrelease} indicates a direct correlation between the magnitude of heat release rate in this zone and the requisite temperature of auto-ignition (T$_{ig}$). It is noteworthy that among the examined mixtures, the 50\%n-heptane-50\%ethanol blend stands out, because it has the highest value of ignition temperature, T$_{ig}$, a phenomenon that can be attributed to its  lowest peak in heat release rate.

\begin {figure} [h!]
\centerline{\includegraphics[width = 192pt]{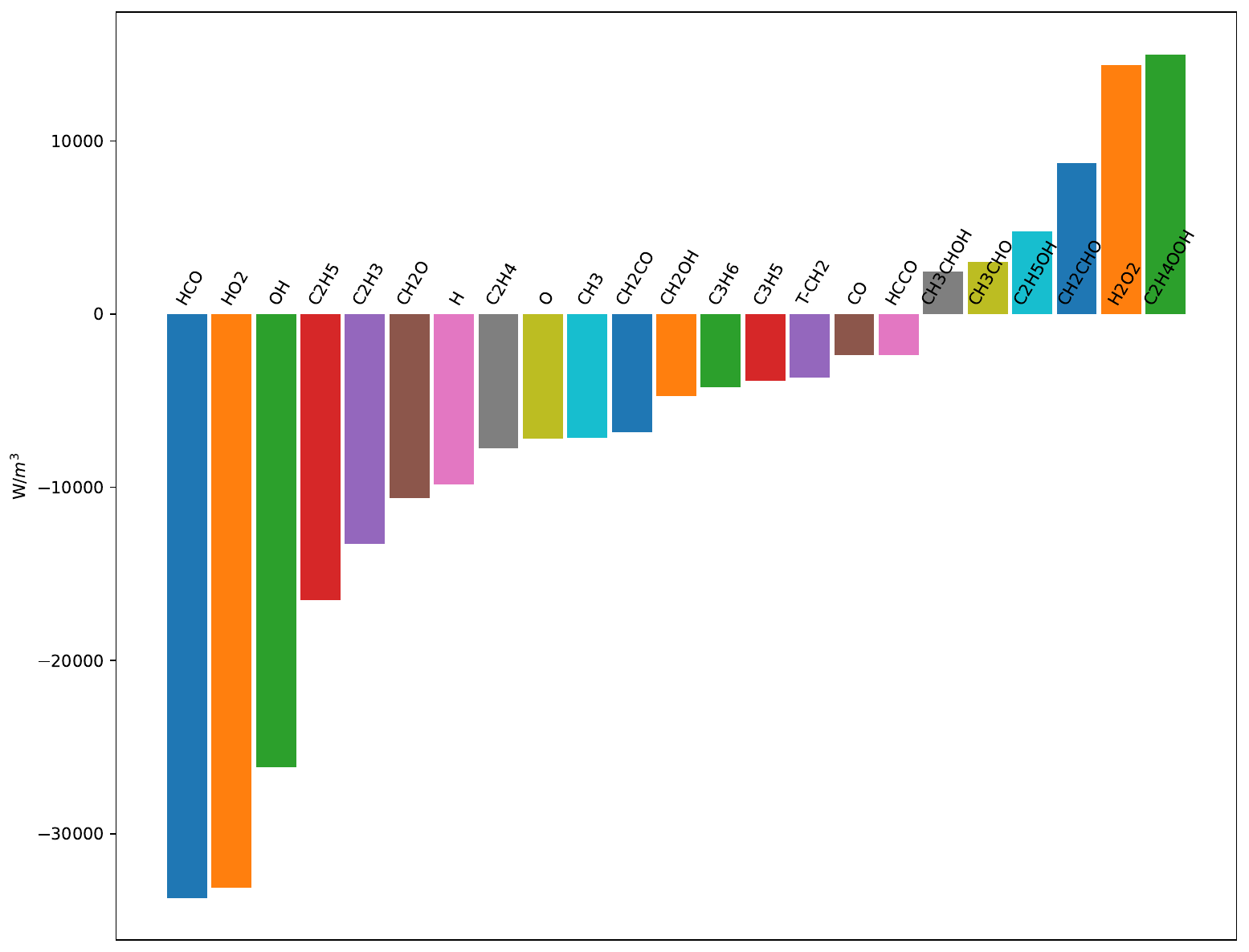}}
\caption{Distribution of heat release rate change from 100\%n-heptane to 90\%n-heptane}
\label{fig:heatreleasedistribution100-90}
\end{figure}

\begin {figure} [h!]
\centerline{\includegraphics[width = 192pt]{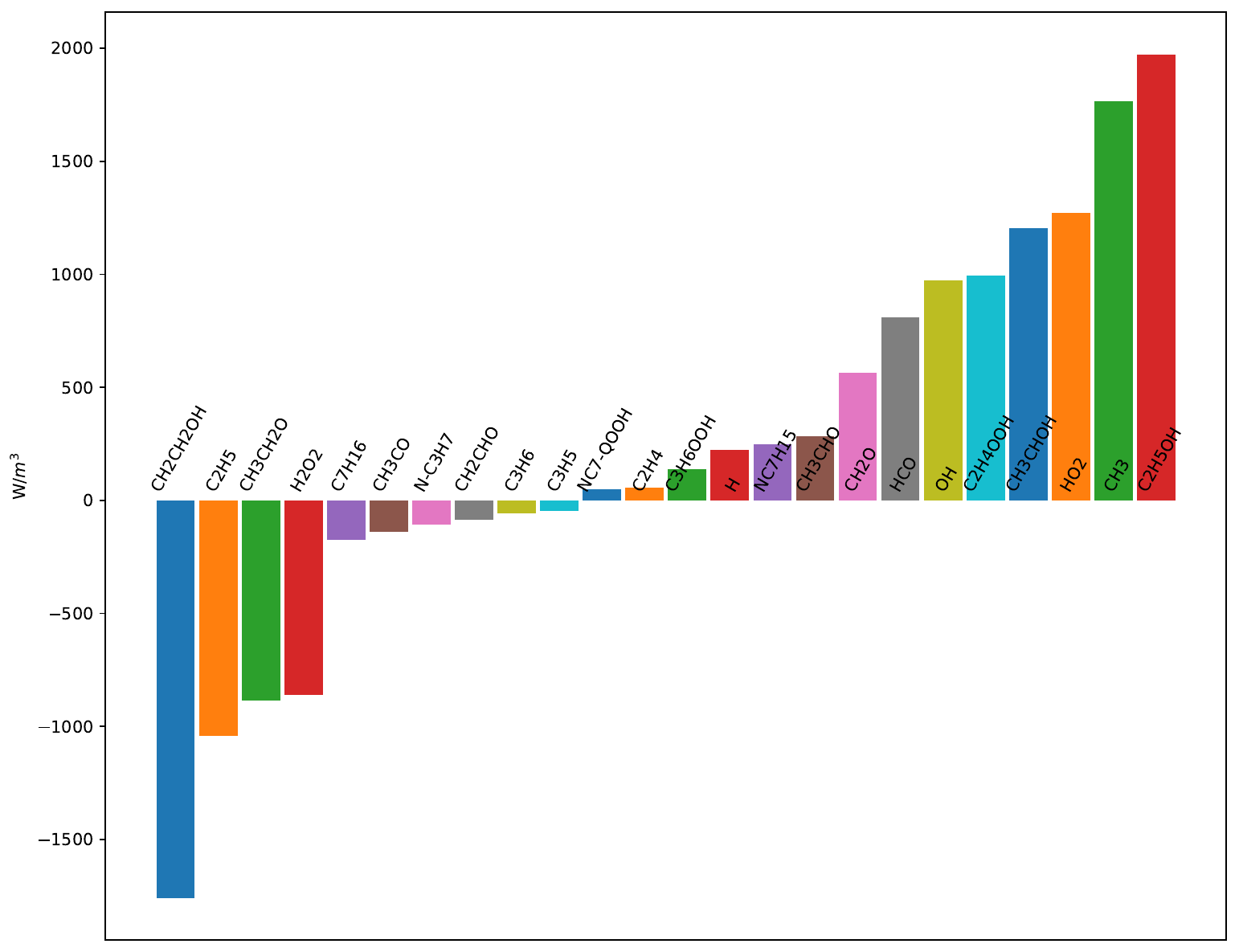}}
\caption{Heat release rate change from 50\% n-heptane to 20\% n-heptane}
\label{fig:heatreleasedistribution50-20}
\end{figure}

Fig.\ \ref {fig:heatrelease} shows that the reduction of the volume fraction of heptane from 100\% to 90\% leads to a significant decrease in heat release rate. Furthermore heat-release-rate-analysis, as depicted in Fig.\ \ref {fig:heatreleasedistribution100-90}, directly attributes this decline of heat release rate predominantly to changes of concentrations of {\hco}, {\hot}, {\oh}, {\cthf}, {\cthtr}, {\chto}. In contrast, the reduction in volume fraction of n-heptane from 50\% to 20\% correlates to an observable increase in heat release rate, as shown in Fig.\ \ref {fig:heatrelease}. Similar results of heat-release-rate-analysis shown in Fig.\ \ref {fig:heatreleasedistribution50-20} attributes the increase of heat release rate to involvement of species such as {\chthr}, {\hot}, {\oh}, {\hco}, {\ethal}, {\chthrchoh}, {\chto}.

\subsection{Key species in n-heptane-dominant mixtures} \addvspace{10pt}

The species contributing to heat release rate change could be placed  in two groups; one that can be considered to  maintain steady-state at a selected location. and the other that does not maintain steady-state.

\begin {figure} [h!]
\centerline{\includegraphics[width = 192pt]{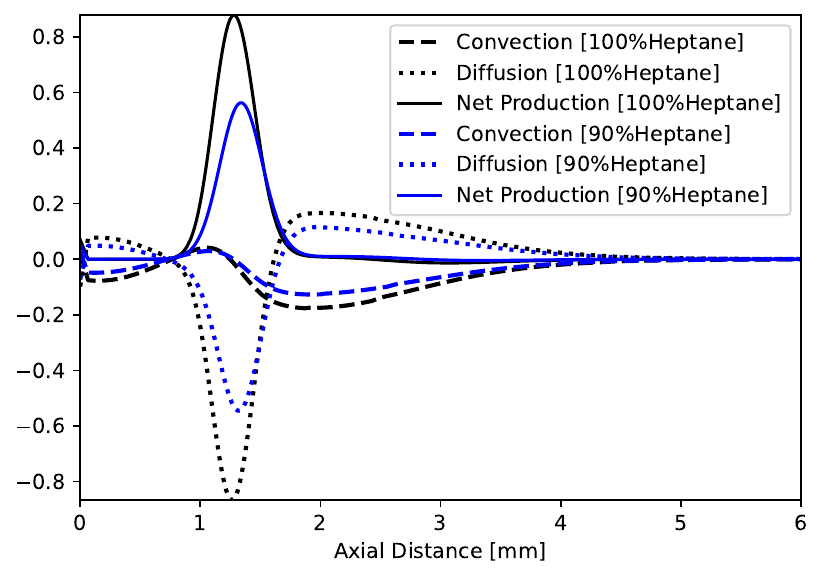}}
\caption{{\chto} Species equation terms}
\label{fig:CH2OEq}
\end{figure}

\begin {figure} [h!]
\centerline{\includegraphics[width = 192pt]{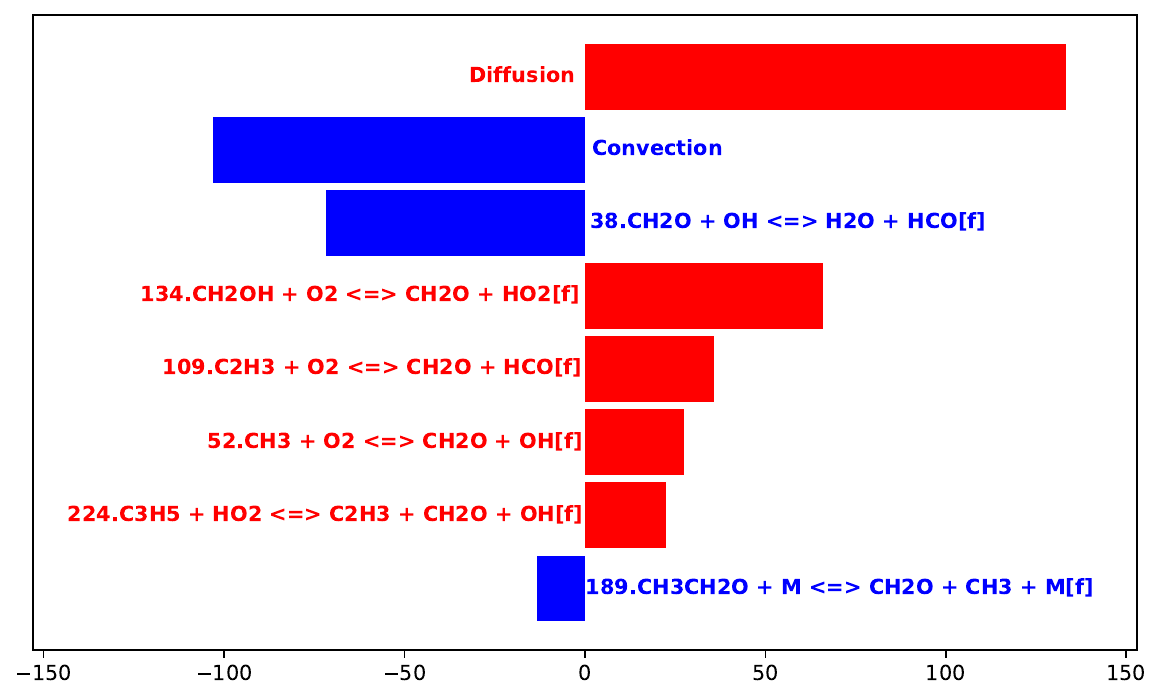}}
\captionsetup{ width=\textwidth}
\caption{Contribution on {\chto} concentration change@4.2mm}
\label{fig:CH2O_high_nc7}
\end{figure}

For species that do not satisfy steady-state approximation, their concentrations are governed by  a species equation that includes diffusion term, convection term and chemical reaction term. Specifically, in Fig.\ \ref{fig:CH2OEq}, for {\chto}, that does not maintain steady-state the peak in the net production  is observed around 1.2{\,}mm, that is located in a zone where the LTC is highly active. Additionally, the positive values of the diffusion term between 2{\,}mm and {\,}mm elucidate the role of the diffusion term in promoting the transport of the {\chto} from the low-temperature zone to the high-temperature zone. Consequently, a decline of activity of LTC in the low-temperature region, caused by additional ethanol, leads to a reduction of diffusion effect, subsequently impacting heat release rate and auto-ignition at the high-temperature region. This effect is also evidenced by analysis of {\chto}  concentration change, depicted in Fig.\ \ref{fig:CH2O_high_nc7}. The observed reduction of {\chto} in the high-temperature region is primarily attributed to a decrease in the value of the diffusion term. Similar behaviors are observed with other species, specifically formaldehyde (\cthfo), ethylene (\ctrhs), and hydrogenperoxide (\htot). The corresponding analyses and plots for these species are shown in the supplemental materials, ranging from Fig.\ \ref{fig:C2H4Eq} to Fig.\ \ref{fig:H2O2_high_nc7}.

For species that are considered to maintain steady-state, their concentrations are predominantly governed by the equilibrium between production and consumption rates at a selected location. Among the primary species shown in Fig.\ \ref{fig:heatreleasedistribution100-90} that leads to heat release rate reduction,  {\hco}, {\hot}, {\oh}, {\cthf} and {\cthtr} are identified to be in the steady-state. This indicates these radicals are produced and consumed at approximately the same rate at the selected location. Consequently, changes in their concentrations at this location are primarily affected by certain elementary reactions, instead of species diffusion or convection. Moreover, these reactions is controlled by variations of other species in the chemical system. These related reactions and species can be elucidated through further analysis of concentration change.

\begin {figure} [h!]
\centerline{ \includegraphics[width = 192pt]{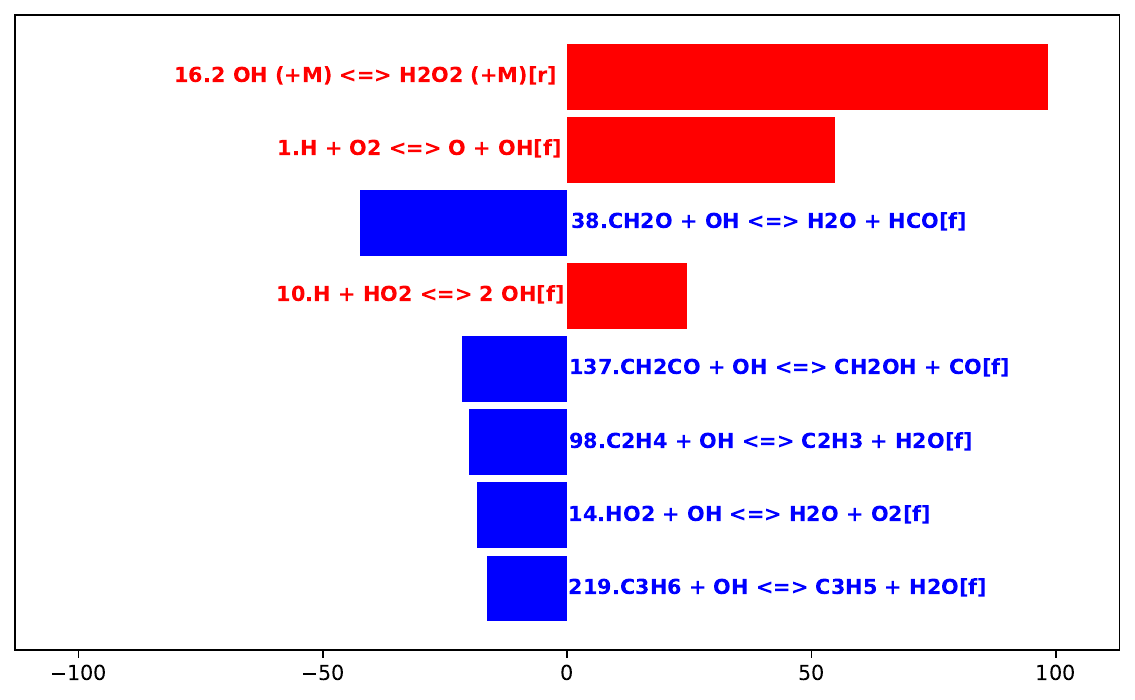}}
\caption{Contribution to {\oh} concentration change@4.2mm}
\label{fig:OH_nc7}
\end{figure}

Fig.\ \ref{fig:OH_nc7} demonstrates that changes in concentration of radical {\oh} at the high-temperature zone are primarily affected by reverse reaction of R16, which is R16r: {\htot} (+M) $\rightarrow$ 2 {\oh} (+M) , involving the decomposition of {\htot}; {\htot} is not in steady state and is diffused from the low-temperature zone. 

\begin {figure} [h!]
\centerline{\includegraphics[width = 192pt]{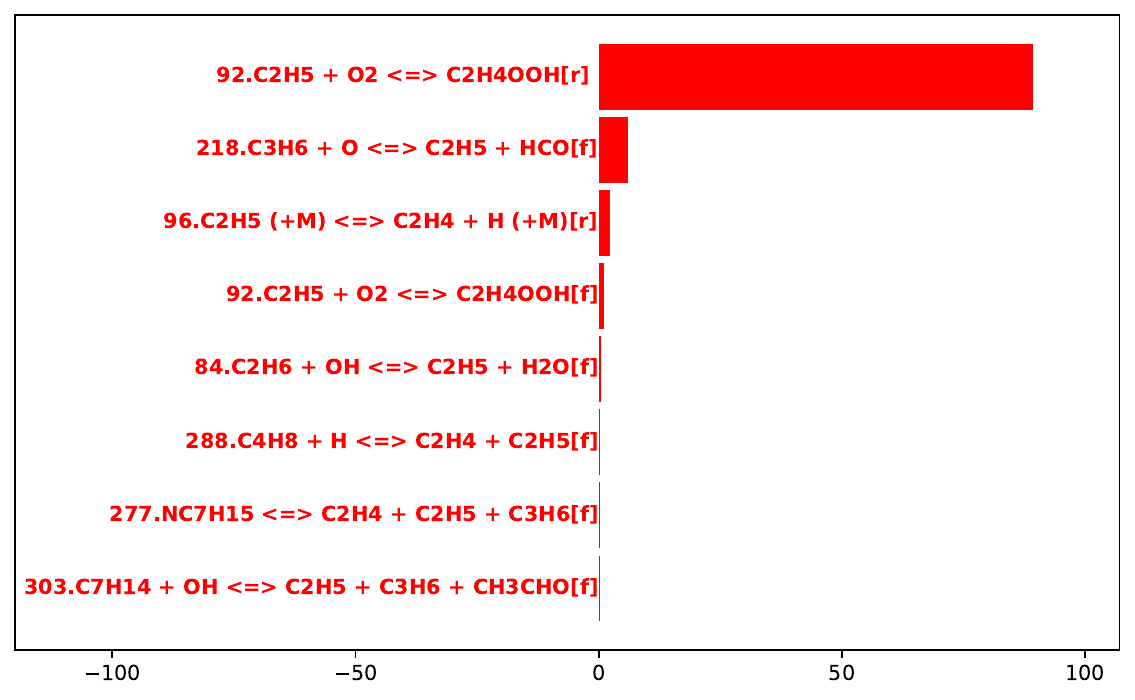}}
\caption{Contribution on {\cthf} concentration change@4.2mm}
\label{fig:C2H5_nc7}
\end{figure}

Further concentration analysis reveals that concentration of {\hot} is predominantly controlled by {\hco}, through R33f: {\hco} + {\ot} ${\rightarrow}$ {\co} + {\hot}. Similarly, {\hco} concentration are primarily controlled by {\chto} and {\oh} via R38f: {\chto} + {\oh} ${\rightarrow}$ {\hto} + {\hco}. Concentration of {\cthtr} is primarily regulated by {\cthfo} by R98f: {\cthfo} + {\oh} ${\rightarrow}$ {\cthtr} + {\hto}. Illustrative plots of these relationships are provided in the supplemental materials, as seen in Fig.\  \ref{fig:HO2_nc7}, \ref{fig:HCO_nc7}, and \ref{fig:C2H3_nc7}. As mentioned before, both {\chto} and {\cthfo} are mainly transported from the low-temperature region via diffusion. 

Fig.\ \ref{fig:C2H5_nc7} reveals that the radical {\cthf} is primarily influenced by R92: {\cthf}+{\ot}
$\rightleftharpoons$ {\cthforooh}. However, it is actually a fast reaction so that the substantial consumption of {\cthf} in the forward reaction is counterbalanced by its production in the reverse reaction. Therefore, greater emphasis should be placed to the second reaction depicted in Fig.\ \ref{fig:C2H5_nc7}, R218f: {\ctrhs} + O $\rightarrow$ {\ctrhs} + {\hco}, which is primarily affected by {\ctrhs}. Notably, {\ctrhs} is not in the steady state at the high-temperature zone and is primarily diffused from the low-temperature region.


In summary, radicals including {\chto}, {\cthfo}, {\ctrhs} and {\htot} predominantly diffused from the low-temperature zone, significantly influencing the heat release rate and auto-ignition process in the high-temperature zone. 

\begin {figure} [h!]
\centerline{\includegraphics[width = 192pt]{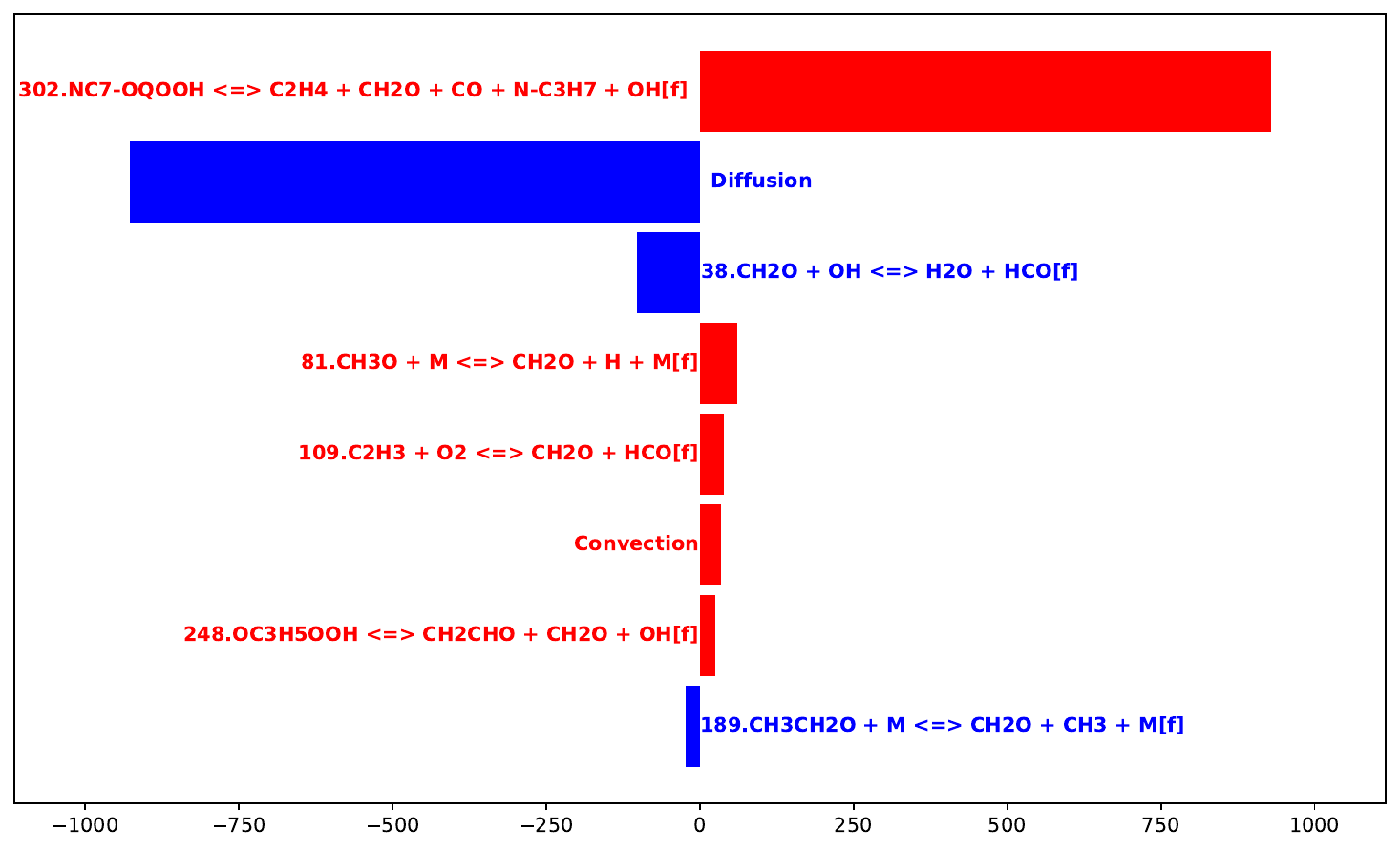}}
\caption{\fussy Contribution on {\chto} concentration change@1.1mm}
\label{fig:CH2O_low_nc7}
\end{figure}

\begin {figure} [h!]
\centerline{\includegraphics[width = 192pt]{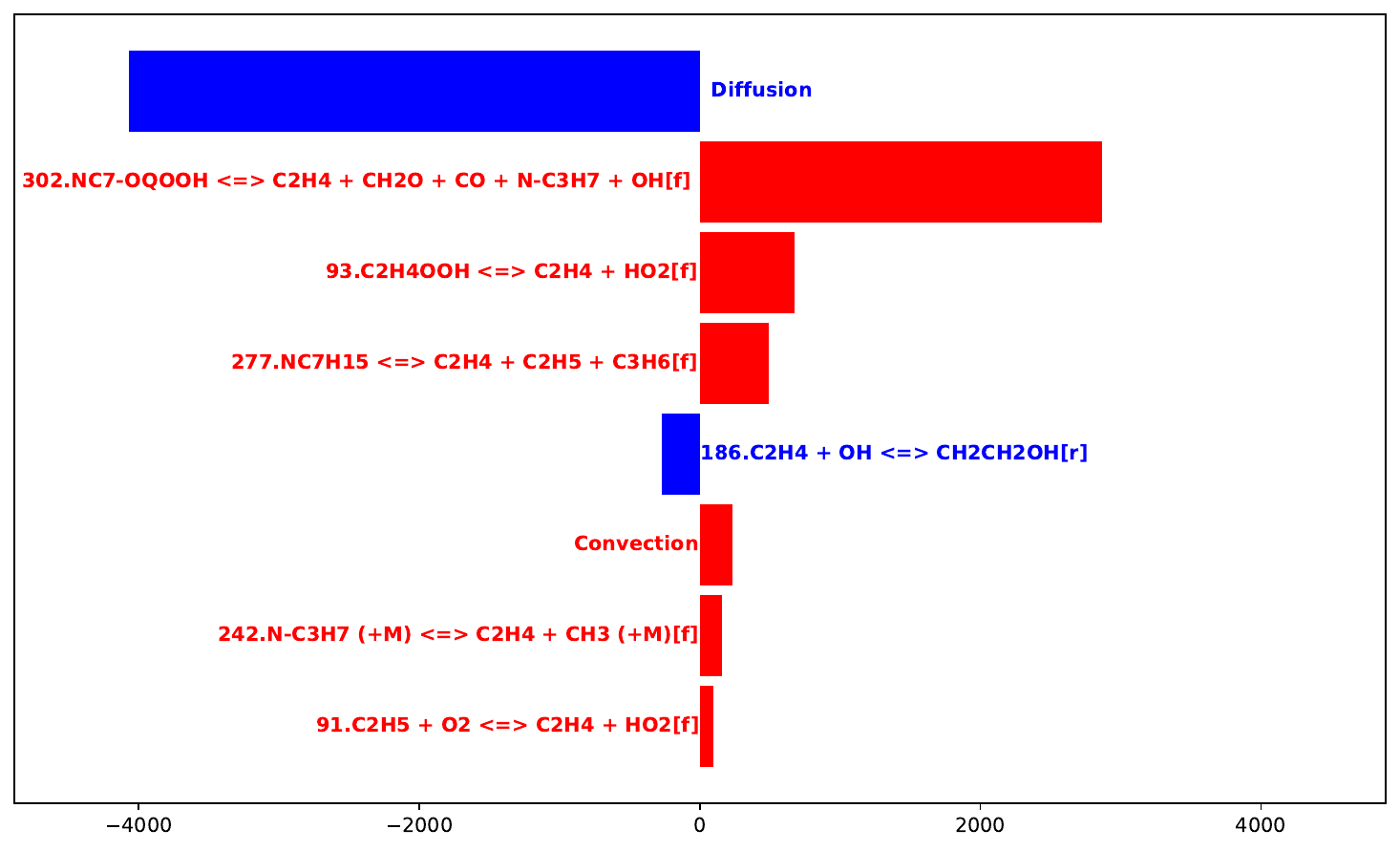}}
\caption{\fussy Contribution on {\cthfo} concentration change@1.1mm}
\label{fig:C2H4_low_nc7}
\end{figure}

This observation underscores the necessity of delineating the production mechanism of these species in the low-temperature zone. As indicated in Fig.\ \ref{fig:CH2O_low_nc7} and Fig.\ \ref{fig:C2H4_low_nc7}, {\chto} and {\cthfo} primarily produced through reaction R302f: NC7OQOOH ${\rightarrow}$ {\cthfo} + {\chto} + {\co} + {\ctrhsvn} + {\oh} at the low-temperature zone.

\begin {figure} [h!]
\centerline{ \includegraphics[width = 192pt]{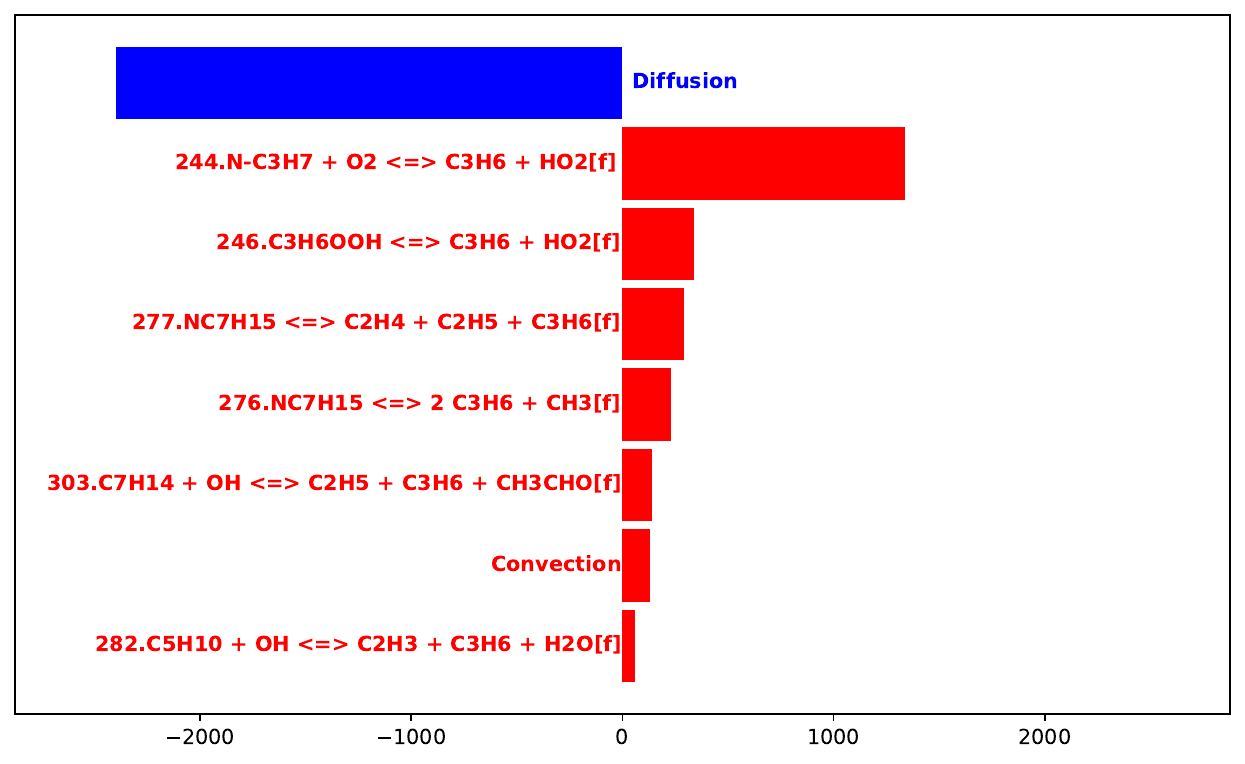}}
\caption{\fussy Contribution on {\ctrhs} concentration change@1.1mm}
\label{fig:C3H6_low_nc7}
\end{figure}

The origins of {\ctrhs} and {\htot} in the low-temperature zone is notably complex. Fig.\ \ref{fig:C3H6_low_nc7} proves that concentration of {\ctrhs} is predominantly regulated by {\ctrhsvn}. Similarly, concentration analysis of {\htot} and {\hot} indicate {\htot} is mainly converted from {\hot}, which, in turn, is affected by {\ctrhsvn}. These analysis results are further detailed in the supplemental materials, specifically in Fig.\ \ref{fig:H2O2_low_nc7} and \ref{fig:HO2_low_nc7}.

\begin {figure} [h!]
\centerline{\includegraphics[width = 192pt]{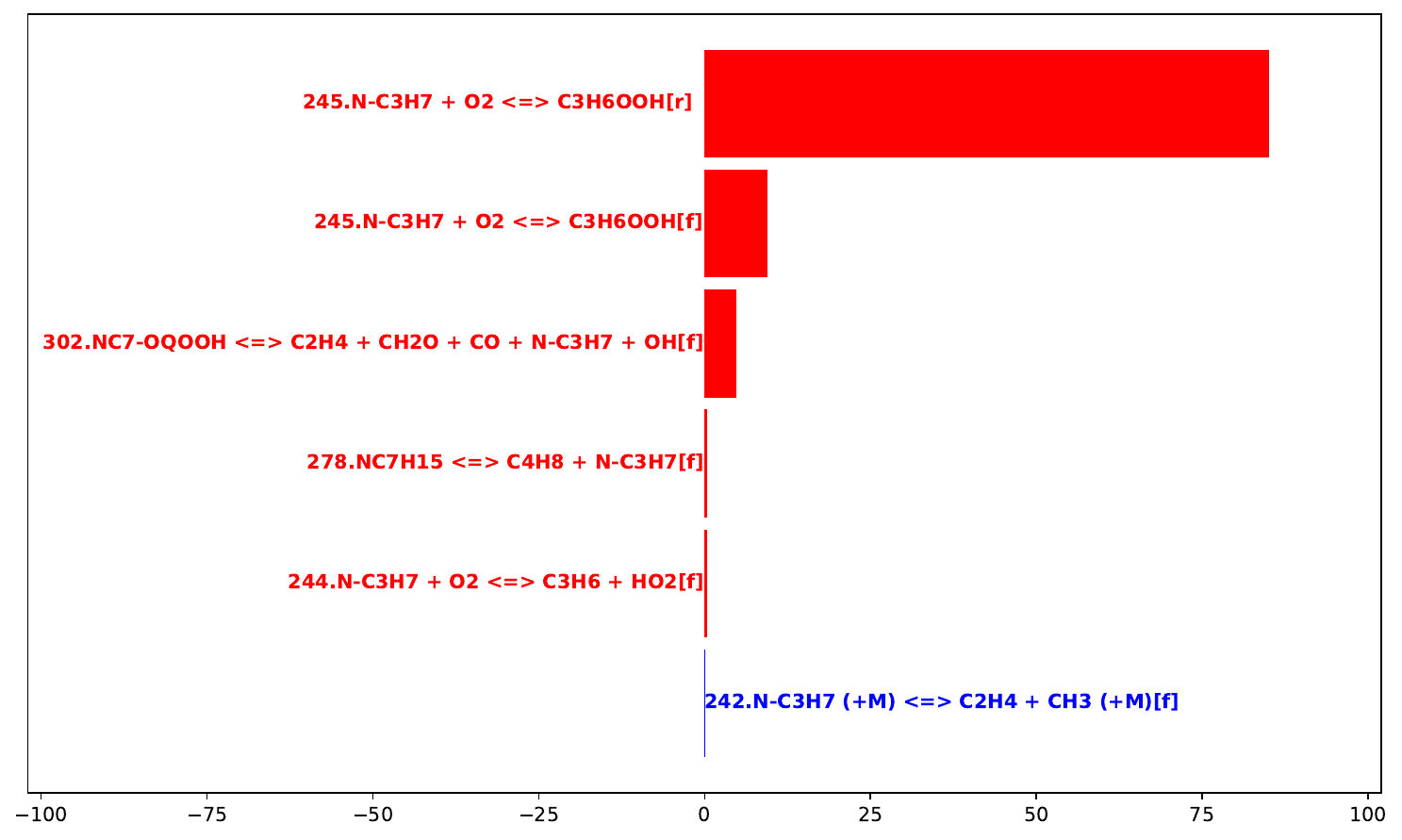}}
\caption{\fussy Contribution on {\ctrhsvn} concentration change@1.1mm}
\label{fig:N-C3H7_low_nc7}
\end{figure}

Given its pivotal role, {\ctrhsvn} emerges as a crucial species impacting both {\htot} and {\ctrhs}. It is primarily produced through reaction R302f, as indicated in Fig.\ \ref{fig:N-C3H7_low_nc7}. Thus, in the low-temperature region, R302f exerts a direct influence on producing {\chto}, {\cthfo}, while indirectly affecting {\ctrhs} and {\hot} via {\ctrhsvn}.

\begin {figure} [h!]
\centerline{\includegraphics[width = 192pt]{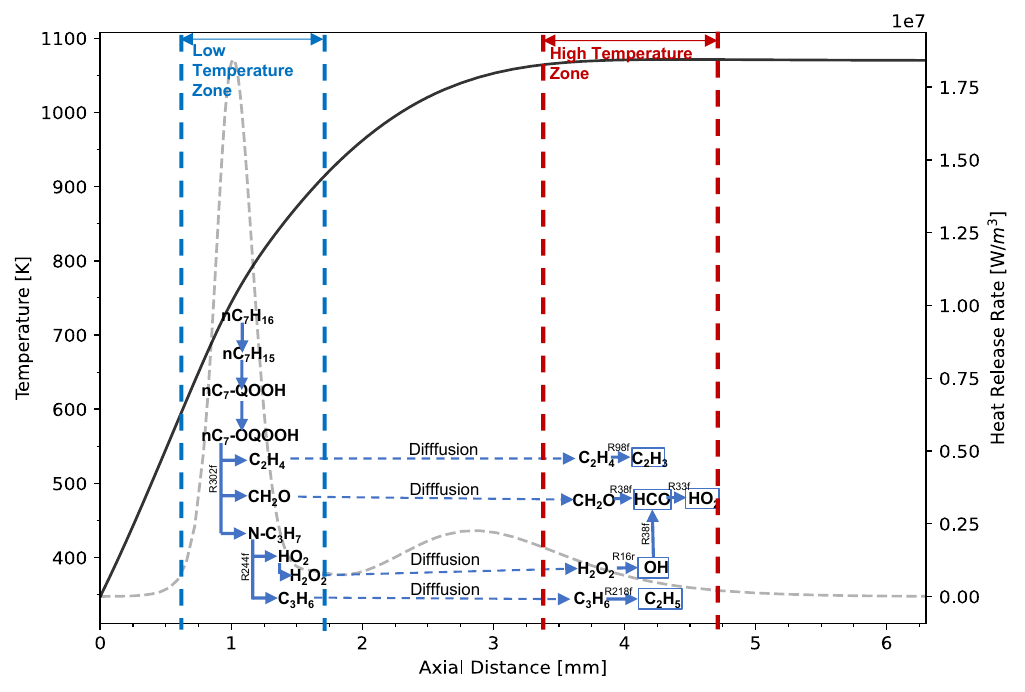}}
\caption{Heptane-dominant mixture overview. In the background, the black solid line represents the temperature profile and the grey dashed line indicates heat release rate }
\label{fig:LSR_sum_nc7}
\end{figure}

Fig.\ \ref{fig:LSR_sum_nc7} provides a comprehensive overview elucidating interplay between the low and high temperature zones in n-heptane-dominant mixtures. The addition of ethanol leads to competition of oxygen, resulting in a decreased of {\ncsvnoqooh} concentration in the low-temperature zone\cite{ji2024experimental}. This reduction in concentration of {\ncsvnoqooh} subsequently diminishes the reaction rate of R302f and concentrations of its products. Ultimately, these effects propagated to the high-temperature zone through species diffusion.

\subsection{key species in ethanol-dominant mixtures} \addvspace{10pt}

\begin {figure} [h!]
\centerline{\includegraphics[width = 192pt]{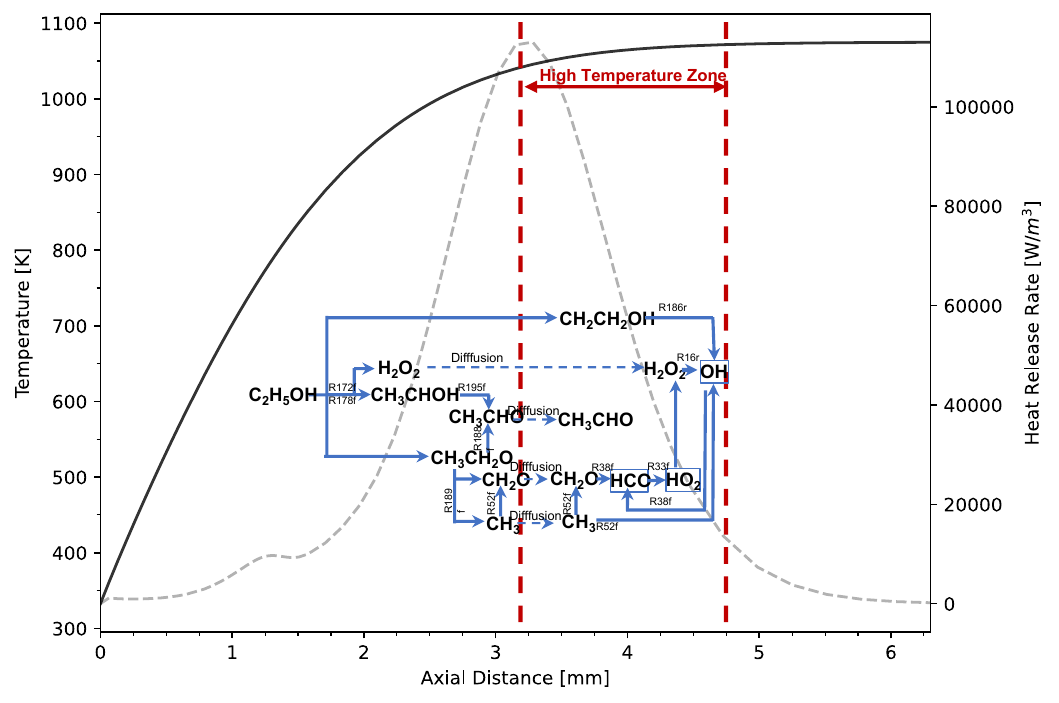}}                                    
\caption{Ethanol-dominant mixture overview. In the background, the black solid line represents the temperature profile and the grey dashed line indicates heat release rate}
\label{fig:LSR_sum_eth}
\end{figure}

It is observed that a decrease of n-heptane's volume fraction from 50\% to 20\% and a corresponding increase in ethanol's volume fraction from 50\% to 80\%, results in an increase in heat release rate in the high-temperature zone. This is primarily due to the formation of three products: {\chthrchoh}, {\chtchtoh} and {\chthrchto}, as shown in Fig \ref{fig:LSR_sum_eth}. These products are formed from ethanol undergoing decomposition through hydrogen abstraction. \cite{SARATHY201440} 

\begin {figure} [h!]
\centerline{\includegraphics[width = 192pt]{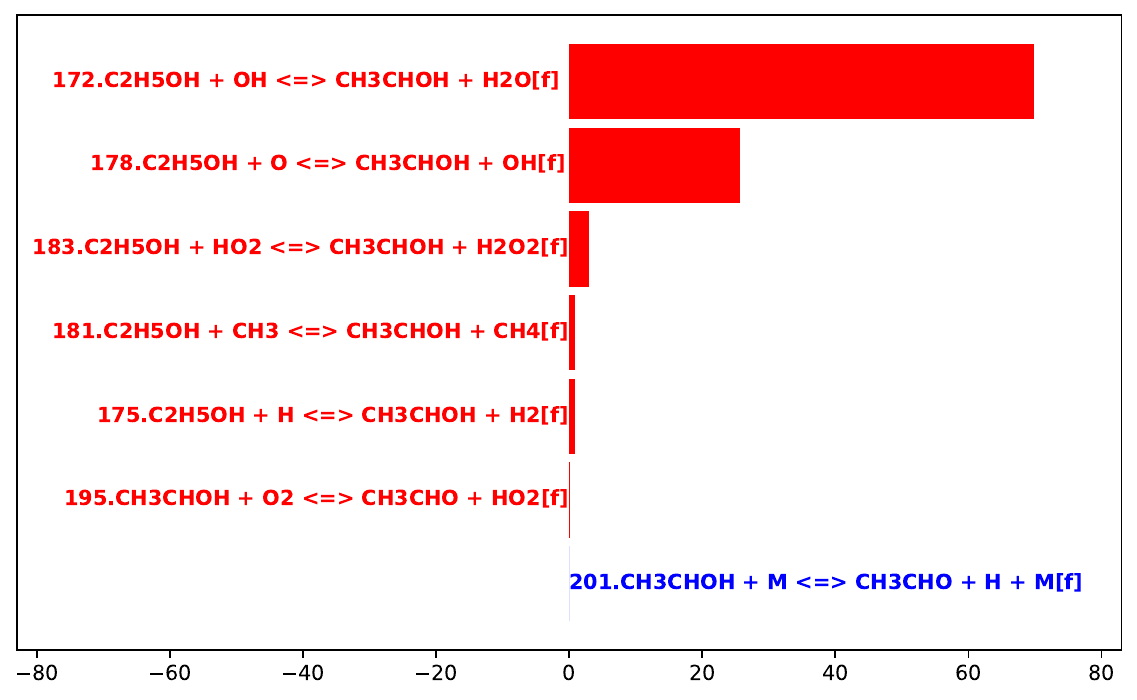}}
\captionsetup{ width=\textwidth}
\caption{Contribution on {\chthrchoh} concentration change@4.2mm}
\label{fig:CH3CHOH}
\end{figure}

\begin{figure}[h!]
    \centering
    \begin{subfigure}[b]{192pt}
        \includegraphics[width=\textwidth]{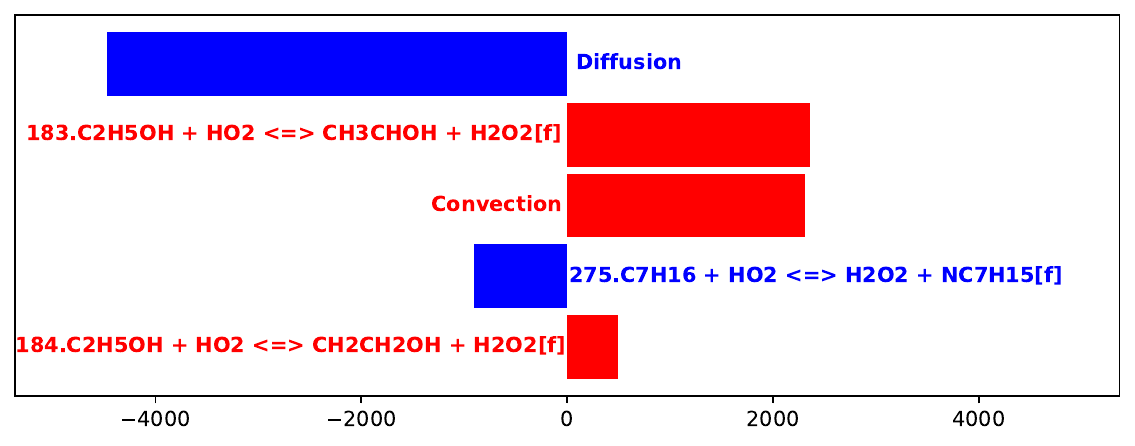}
         \caption{At production peak (2.2mm)}
        \label{fig:H2O2_low}
    \end{subfigure}
    \vspace{0em} 
    \begin{subfigure}[b]{192pt}
        \includegraphics[width=\textwidth]{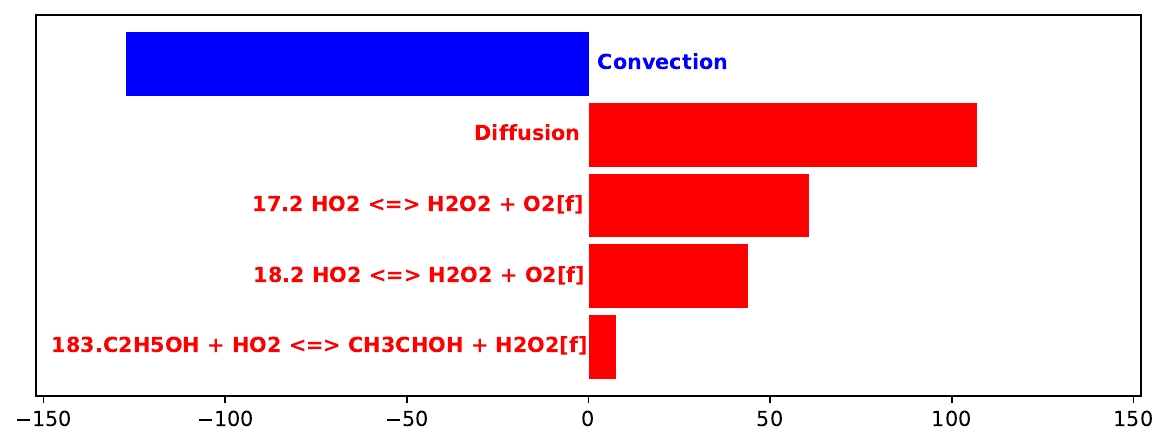}
         \caption{At the high-temperature zone (4.2mm)}
        \label{fig:H2O2_high}
    \end{subfigure}
    \caption{Contribution on {\htot} concentration change}
\end{figure}

\begin {figure} [h!]
\centerline{\includegraphics[width = 192pt]{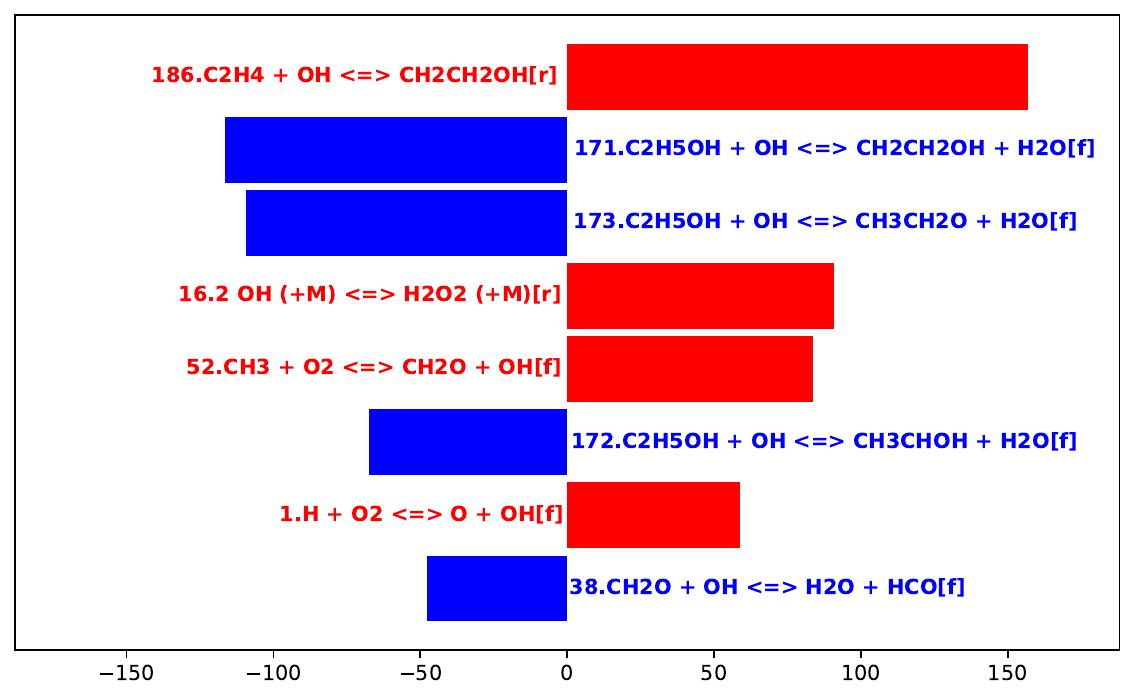}}
\caption{Contribution on {\oh} concentration change @4.2mm}
\label{fig:OH}
\end{figure}

The production of {\chthrchoh}, mainly from reactions R172f: {\ethal} + {\oh} $\rightarrow$ {\chthrchoh} + {\hto} and R178f: {\ethal} + O $\rightarrow$ {\chthrchoh} + {\oh}, plays a significant role in increasing heat release rate, as confirmed in Fig.\ \ref {fig:CH3CHOH}. Notably, one pathway to form {\chthrchoh}, via R183f:  {\ethal} + {\htot} $\rightarrow$ {\chthrchoh} + {\htot}, is accompanied by a significant production of {\htot}, as depicted in Fig.\ \ref{fig:H2O2_high},  with the peak production around 2.0mm-2.2mm, shown in Fig.\ \ref{fig:H2O2_low}. It is subsequently diffused to the high-temperature region, contributing to the increase of {\oh} concentration through R16r , {\htot} (+M) $\rightarrow$ 2 {\oh} (+M),  corroborated by Fig.\ \ref{fig:OH}. Additionally, the reverse reaction of R186, involving the decomposition of {\chtchtoh}, contributes to the elevation of {\oh} concentration, as indicated in Fig.\ \ref{fig:OH}.

\begin{figure}[h!]
    \centering
    \begin{subfigure}[b]{192pt}
        \includegraphics[width=\textwidth]{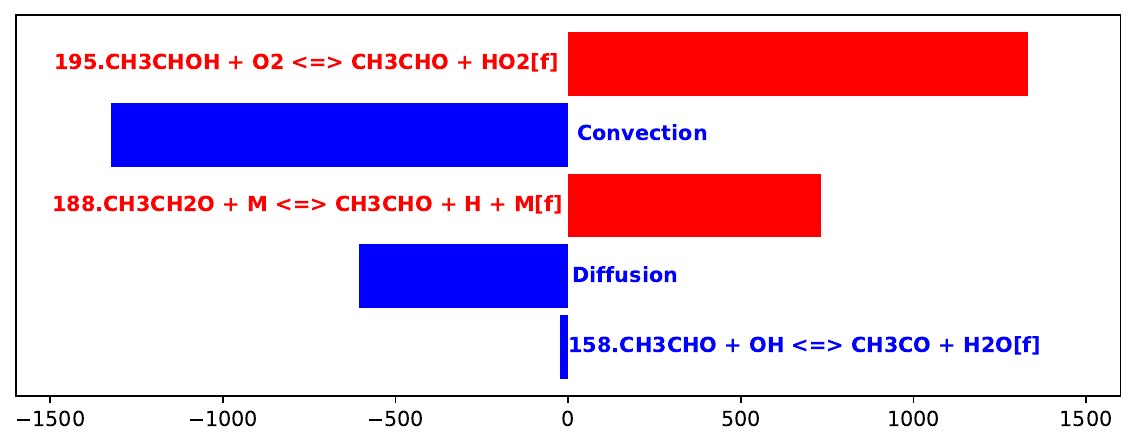}
         \caption{ At the production peak (3.0mm)}
        \label{fig:CH3CHO_low}
    \end{subfigure}
    \vspace{0em} 
    \begin{subfigure}[b]{192pt}
        \includegraphics[width=\textwidth]{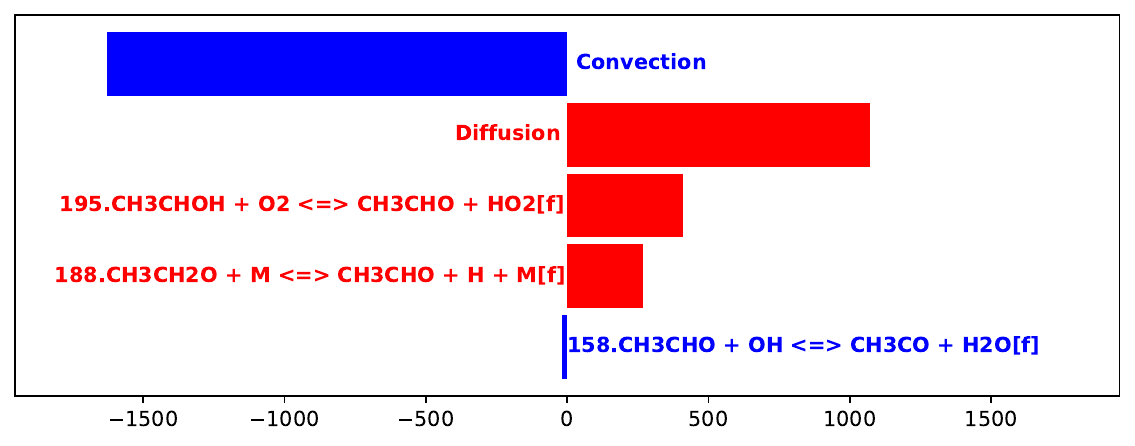}
         \caption{At the high-temperature zone (4.2mm)}
        \label{fig:CH3CHO_high}
    \end{subfigure}
    \caption{Contribution on {\chthrcho} concentration change}
\end{figure}

Furthermore, Fig.\ \ref{fig:CH3CHO_low} demonstrates that the elevated levels of {\chthrchoh} (through R195f) and {\chthrchto} (through R188f) lead to an increase in {\chthrcho}. Furthermore, Fig.\ \ref{fig:CH3CHO_high}  confirms that some of the  of increase of {\chthrcho} is by diffusion from its production peak region to the high-temperature region, thereby promoting heat release rate.

\begin {figure} [h!]
\centerline{ \includegraphics[width = 192pt]{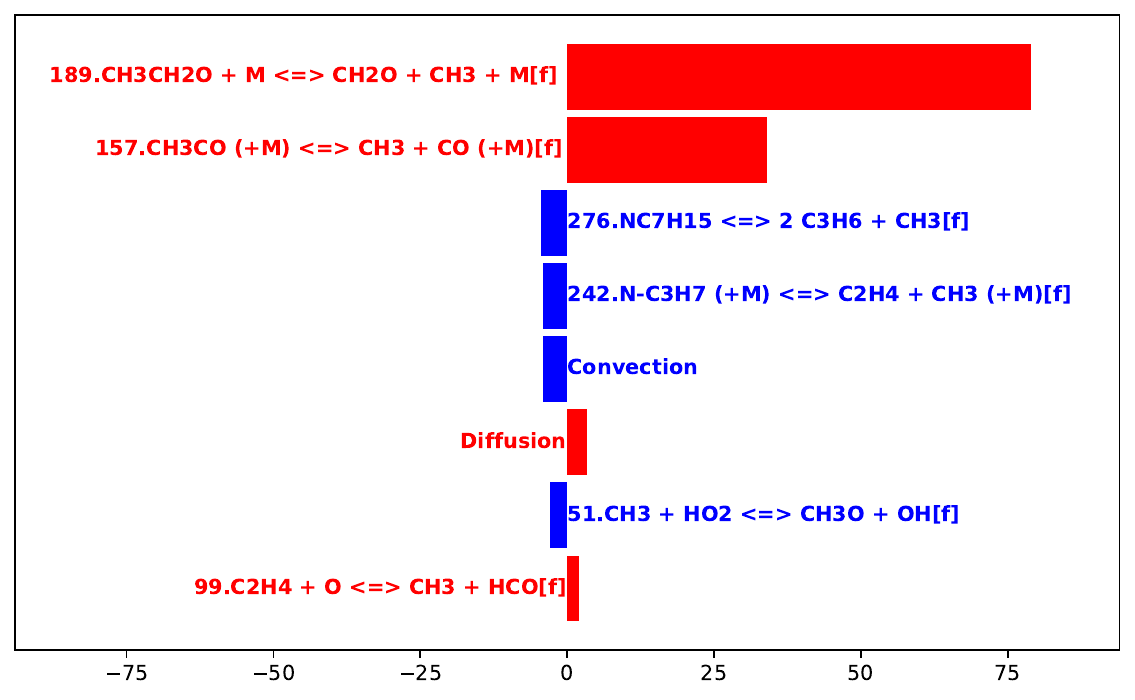}}
\captionsetup{ width=\textwidth}
\caption{Contribution on {\chthr} concentration change@4.2mm}
\label{fig:CH3}
\end{figure}

As shown in Fig.\ \ref{fig:CH3}, the products {\chthr} and  {\chto} from decomposition of {\chthrchto} through R189f and from the step {\chthrchto} +M$\rightarrow$ {\chto} + {\chthr} +M, diffuse to the high-temperature region. This process is augmented by R52f: {\chthr} + {\ot} $\rightarrow$ {\chto} + {\oh}, as illustrated in Fig \ref{fig:CH2O}, wherein {\chthr} reacts with {\ot}, further elevating the concentration of {\chto}. The increase in {\chto} and {\oh} leads to a rise in {\hco} and as a consequence  to an enhanced concentration of {\hot} in the high-temperature region. The concentration change analysis related to  {\chto}, {\hco} and {\hot}  are provided in Fig.\ \ref{fig:CH2O}, Fig.\ \ref{fig:HCO} and Fig.\ \ref{fig:HO2} in supplemental materials.

The reaction pathway including {\chto}, {\hco} and {\hot} exhibits similarity in both n-heptane-dominant and ethanol-dominant mixtures. However, the concentration of {\chto} in n-heptane-dominant mixtures is predominantly influenced by the low-temperature chemistry of n-heptane, whereas in ethanol-dominant mixtures, {\chto} is significantly affected, directly and indirectly, by {\chthrchto}, through ethanol's hydrogen abstraction reaction, R189f.

Additionally, the increase in {\hot} partially results in heightened level of {\htot} at the high-temperature region, as evidenced in Fig.\ \ref{fig:H2O2_high}. 

Fig.\ \ref{fig:LSR_sum_eth} succinctly summarizes the reaction pathways and diffusive effects associated with ethanol-dominant mixtures. The observed increase in heat release rate, as ethanol volume fraction increased from 50\% to 80\%, is primarily attributable to radicals and reactions associated with ethanol's chemistry. In these mixtures, the influence of the inhibition of LTC is considered to be negligible.

\subsection{Summary}
The demonstrated example specifically concerns combustion of n-heptane and ethanol mixtures in a counterflow flame at low strain rate. Simulation results indicate a correlation between heat release rate and auto-ignition temperature. Further quantitative analyses, utilizing the proposed method, reveal that changes in the heat release rate in both n-heptane and ethanol dominant mixtures are associated with the concentration change of certain species in the high-temperature zone.

In n-heptane-dominant mixtures, the investigation reveals that steady-state species at the high-temperature zone, including hydroxyl radicals (\oh) and hydroperoxyl radicals (\hot), maintain the equilibrium between production and consumption rate, directly affecting heat release rate. Non-steady-state species at the high-temperature zone, such as \chto\ and \cthfo\  play significant roles in the auto-ignition process, primarily due to their involvement in LTC at the low-temperature zone and subsequent species diffusion to the high-temperature region. 

For ethanol-dominant mixtures, the study highlights the observed increase in heat release rate with the increase in ethanol's volume fraction, attributing this elevation to the decomposition of ethanol into major products including \chthrchoh, \chtchtoh\ and \chthrchto\ radicals. The production of these species, particularly through hydrogen abstraction reactions, is identified as the key pathway driving the observed increase in heat release rate. 

Using the proposed method, we identified the key species involved in diffusion and demonstrate how the diffusion of these species bridges the low and high temperature zones in n-heptane-dominant mixtures. Additionally, the analysis results indicates, in the ethanol dominant mixtures, chemical kinetics are notably unaffected by n-heptane's LTC, highlighting the distinctive chemical pathways of ethanol and the influence of fuel composition on heat release rate and auto-ignition.
\section{Conclusion and outlook}
This work proposed a method of analysis to reveal the relationship between change in heat release rate and variations of species, elucidating interaction among related species and the potential influence of species transport across different temperature zones in reactive field.

For demonstration, the study investigated the auto-ignition temperature of n-heptane and ethanol mixtures in a counterflow flame configuration under low strain rates. The analysis results indicate that this method effectively quantifies and compares the influence of chemical kinetics and species diffusion effects, detailing the complex interactions of critical species that influence the heat release rate.

Future research directions may include applying the method of analysis under varying strain rates and extending to other one-dimensional flame configurations. Additionally, given the complexity of preparing the overview figures for various mixtures, future efforts will focus on streamlining the analysis process through enhancing code automation.

\appendix
\section{Concentration Analysis Results}
\subsection{n-Heptane Dominant Mixtures}
\begin {figure} [H]
\centerline{\includegraphics[width = 192pt]{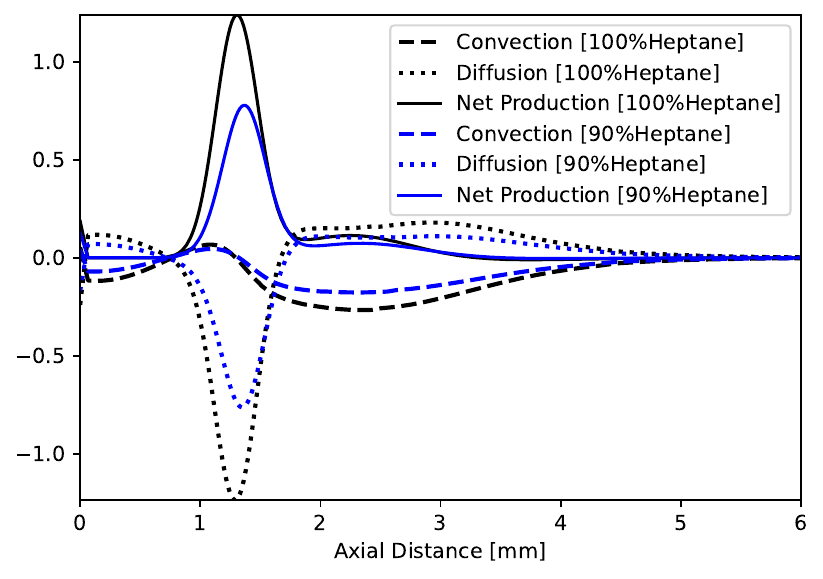}}
\caption{{\cthfo} Species equation terms}
\label{fig:C2H4Eq}
\end{figure}

\begin {figure} [H]
\centerline{\includegraphics[width = 192pt]{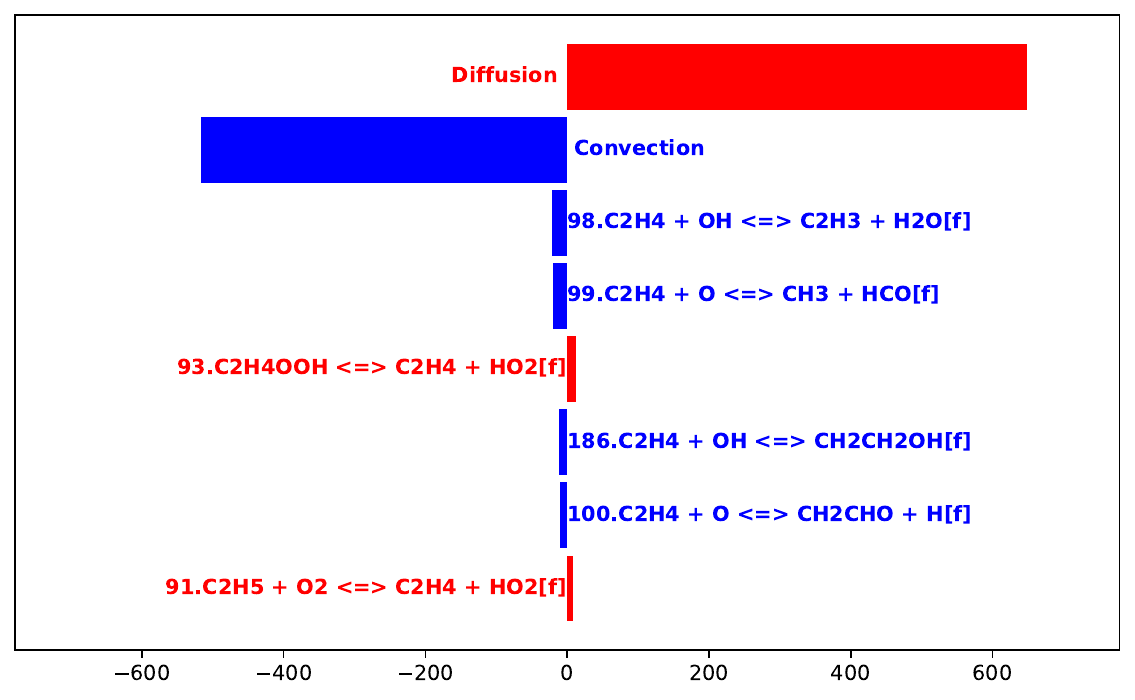}}
\captionsetup{ width=\textwidth}
\caption{Contribution on {\cthfo} concentration change @4.2mm}
\label{fig:C2H4_high_nc7}
\end{figure}

\begin {figure} [H]
\centerline{\includegraphics[width = 192pt]{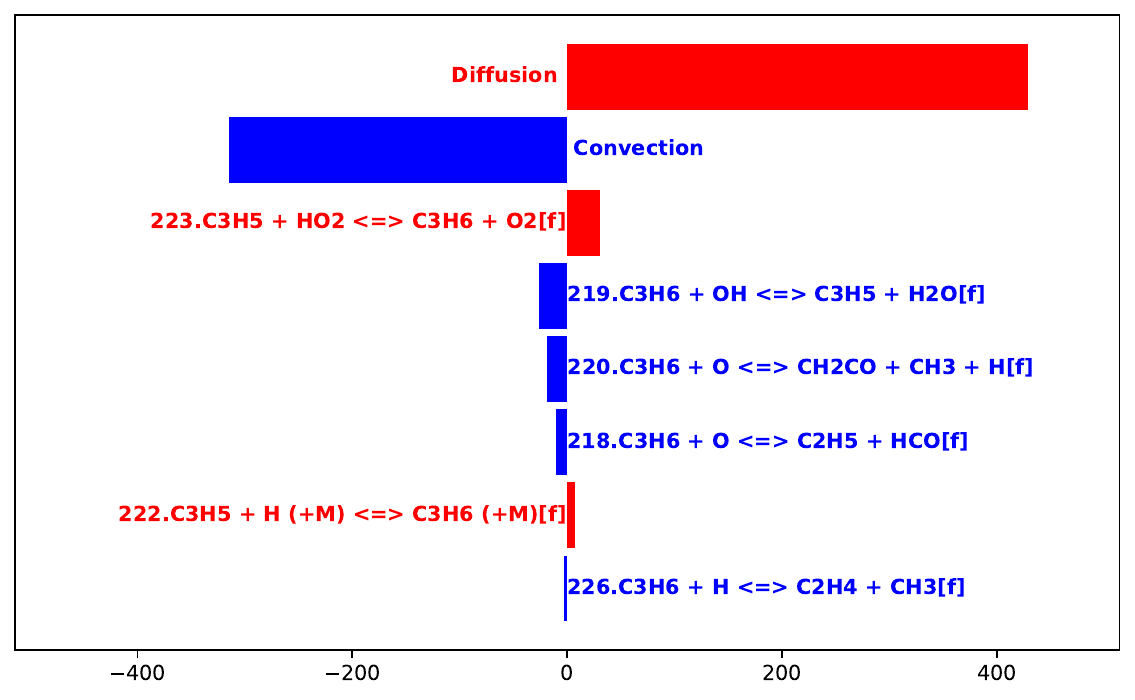}}
\captionsetup{ width=\textwidth}
\caption{Contribution on {\ctrhs} concentration change @4.2mm}
\label{fig:C3H6_high_nc7}
\end{figure}

\begin {figure} [h!]
\centerline{\includegraphics[width = 192pt]{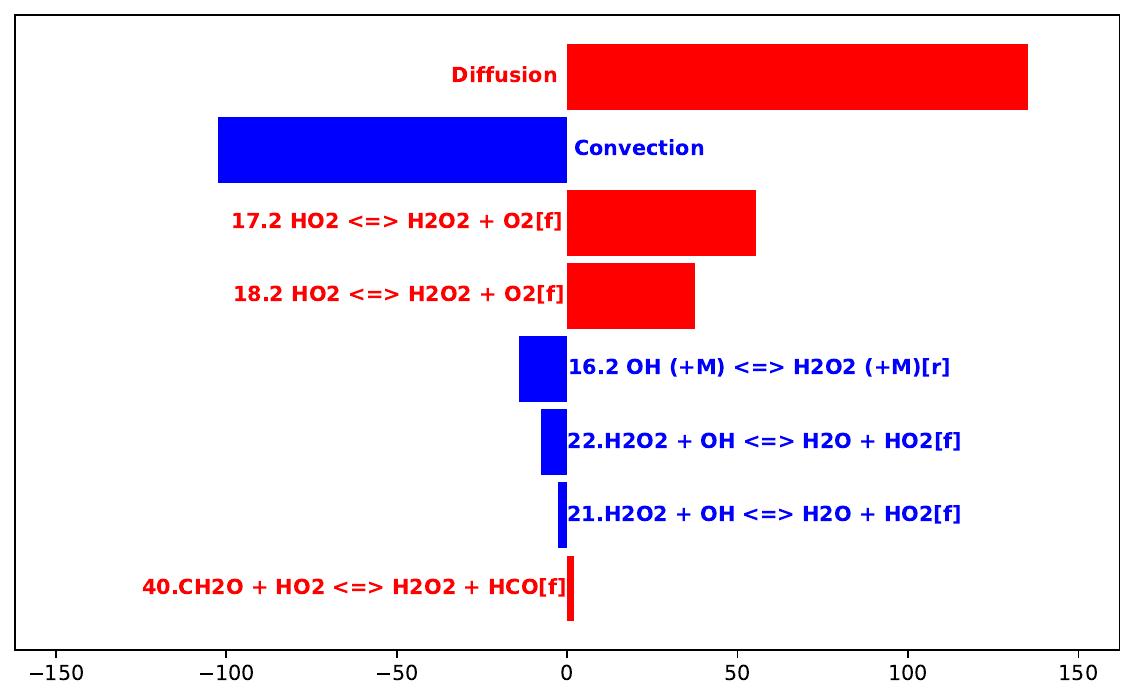}}
\captionsetup{ width=\textwidth}
\caption{Contribution on {\htot} concentration change @4.2mm}
\label{fig:H2O2_high_nc7}
\end{figure}

\begin {figure} [H]
\centerline{\includegraphics[width = 192pt]{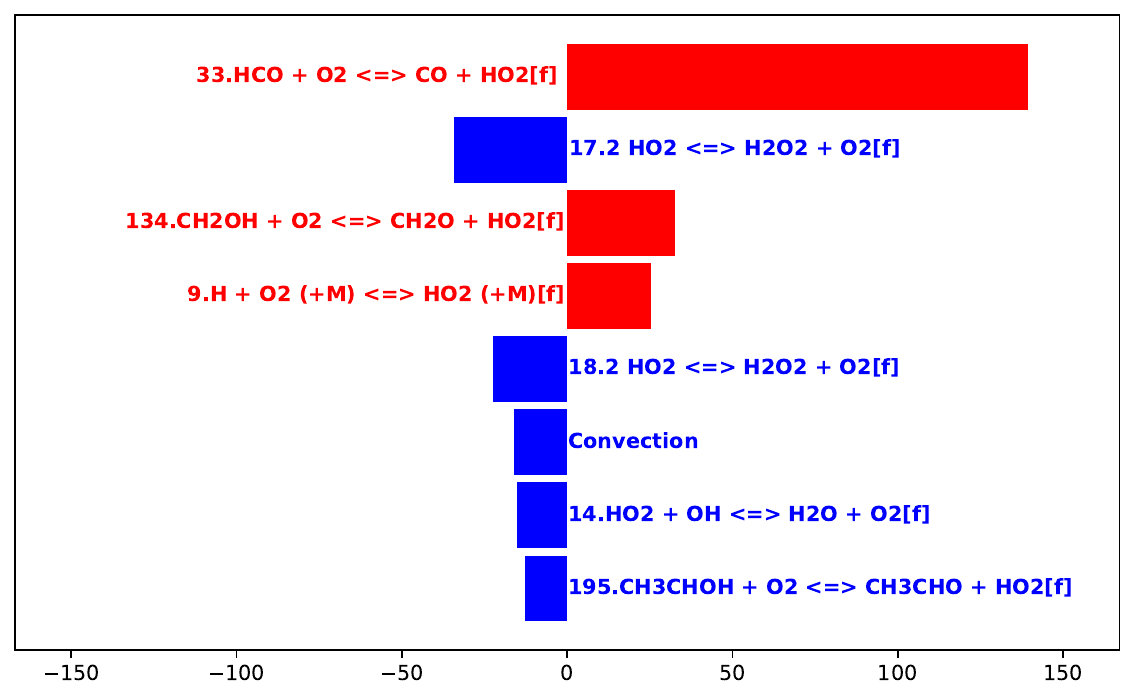}}
\captionsetup{ width=\textwidth}
\caption{Contribution on {\hot} concentration change @4.2mm}
\label{fig:HO2_nc7}
\end{figure}

\begin {figure} [H]
\centerline{\includegraphics[width = 192pt]{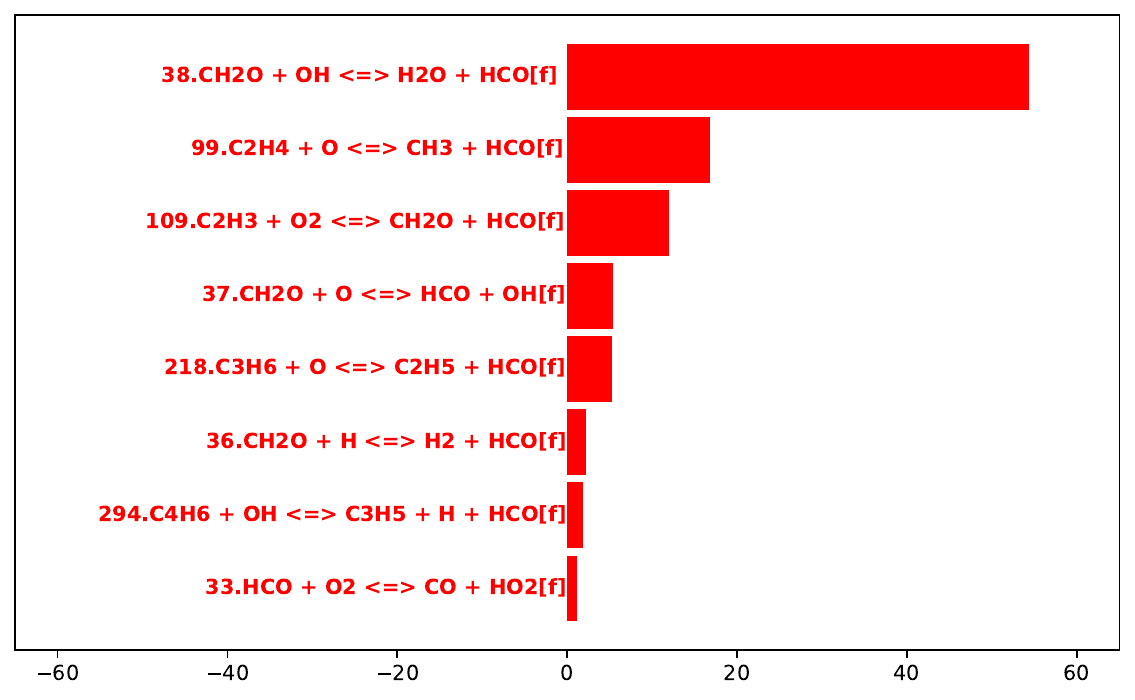}}
\captionsetup{ width=\textwidth}
\caption{Contribution on {\hco} concentration change @4.2mm}
\label{fig:HCO_nc7}
\end{figure}

\begin {figure} [H]
\centerline{\includegraphics[width = 192pt]{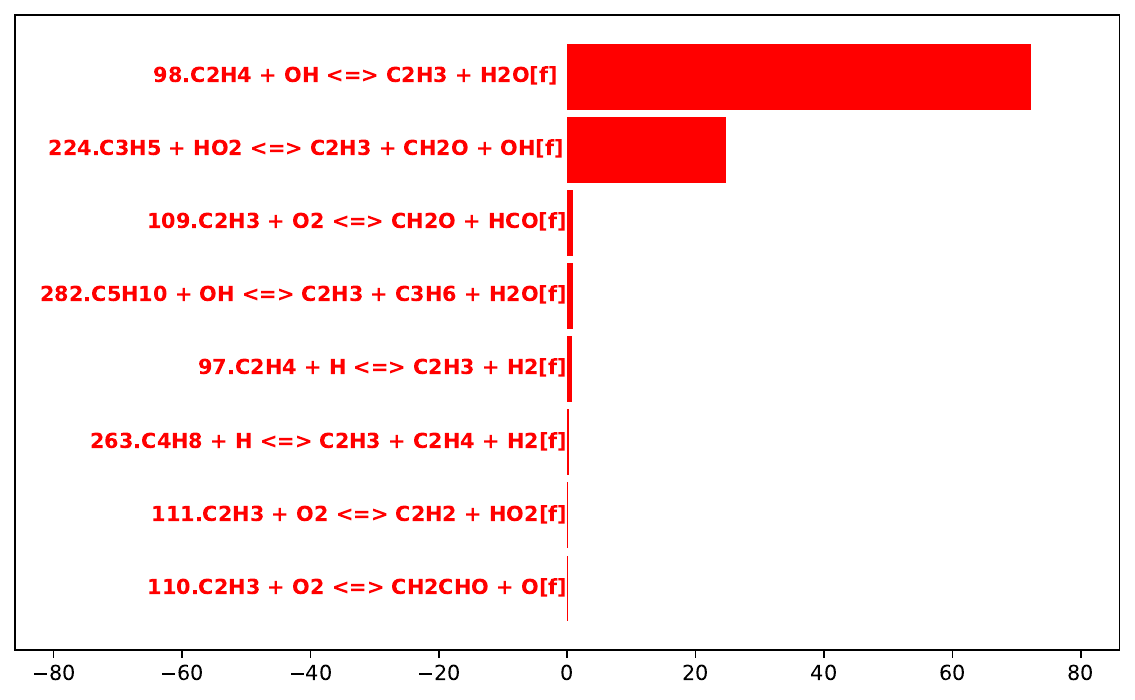}}
\captionsetup{ width=\textwidth}
\caption{Contribution on {\cthtr} concentration change @4.2mm}
\label{fig:C2H3_nc7}
\end{figure}

\begin {figure} [H]
\centerline{\includegraphics[width = 192pt]{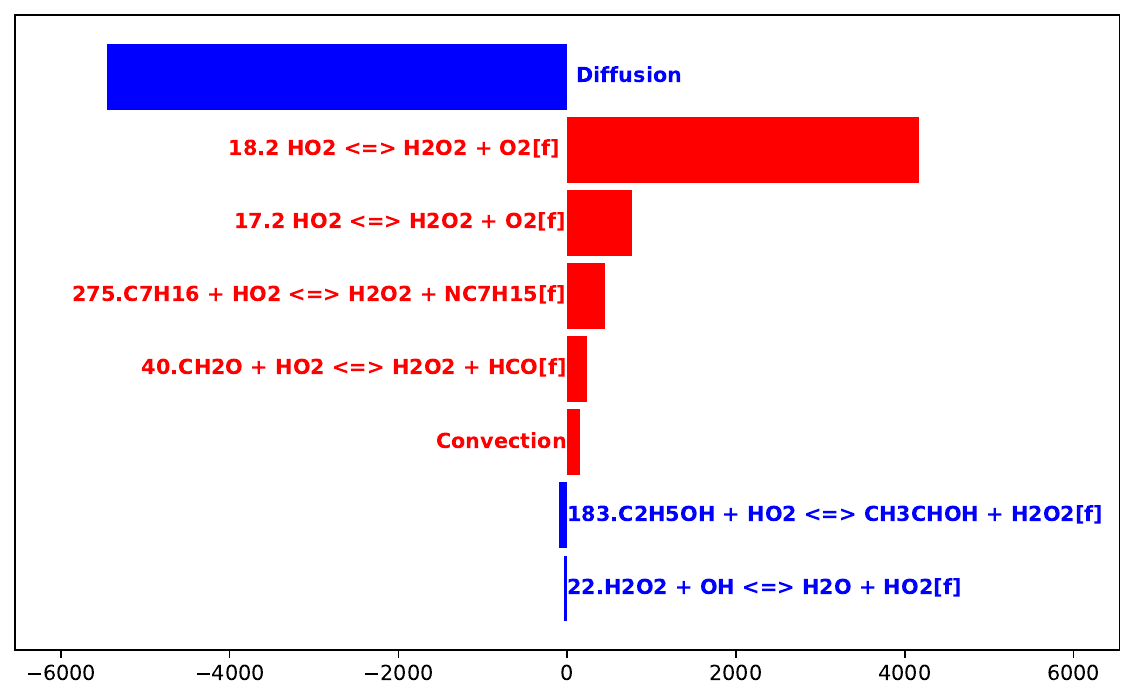}}
\captionsetup{ width=\textwidth}
\caption{Contribution on {\htot} concentration change @1.1mm}
\label{fig:H2O2_low_nc7}
\end{figure}

\begin {figure} [H]
\centerline{\includegraphics[width = 192pt]{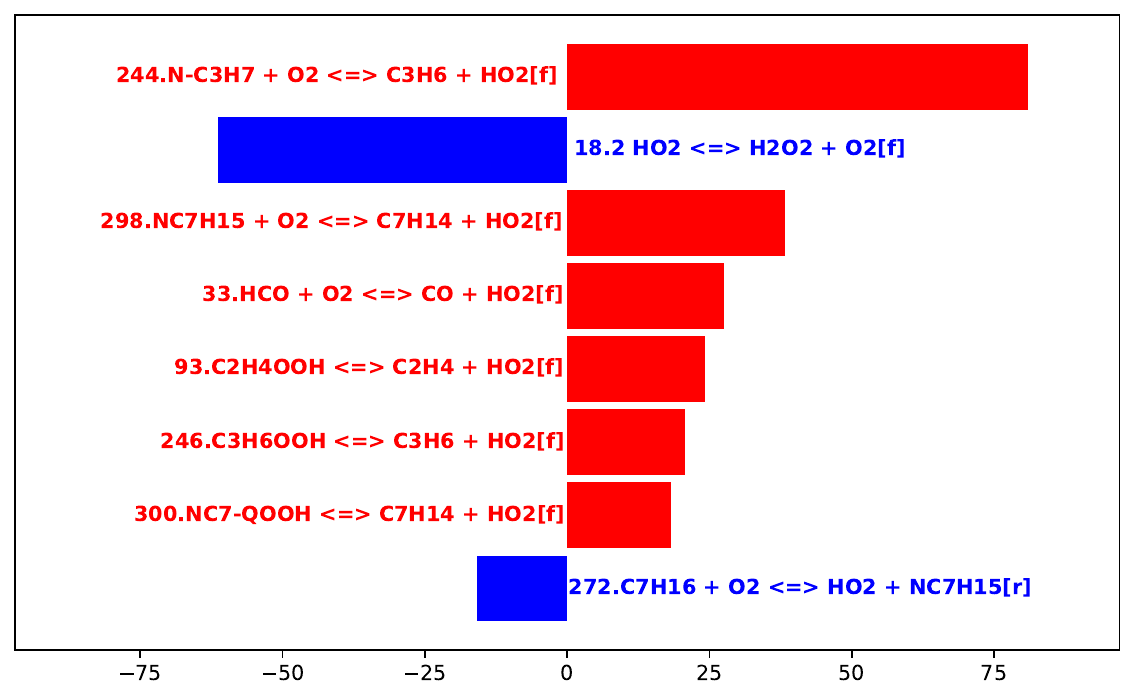}}
\captionsetup{ width=\textwidth}
\caption{Contribution on {\hot} concentration change @1.1mm}
\label{fig:HO2_low_nc7}
\end{figure}

\subsection{Ethanol Dominant Mixtures}
\begin {figure} [H]
\centerline{\includegraphics[width = 192pt]{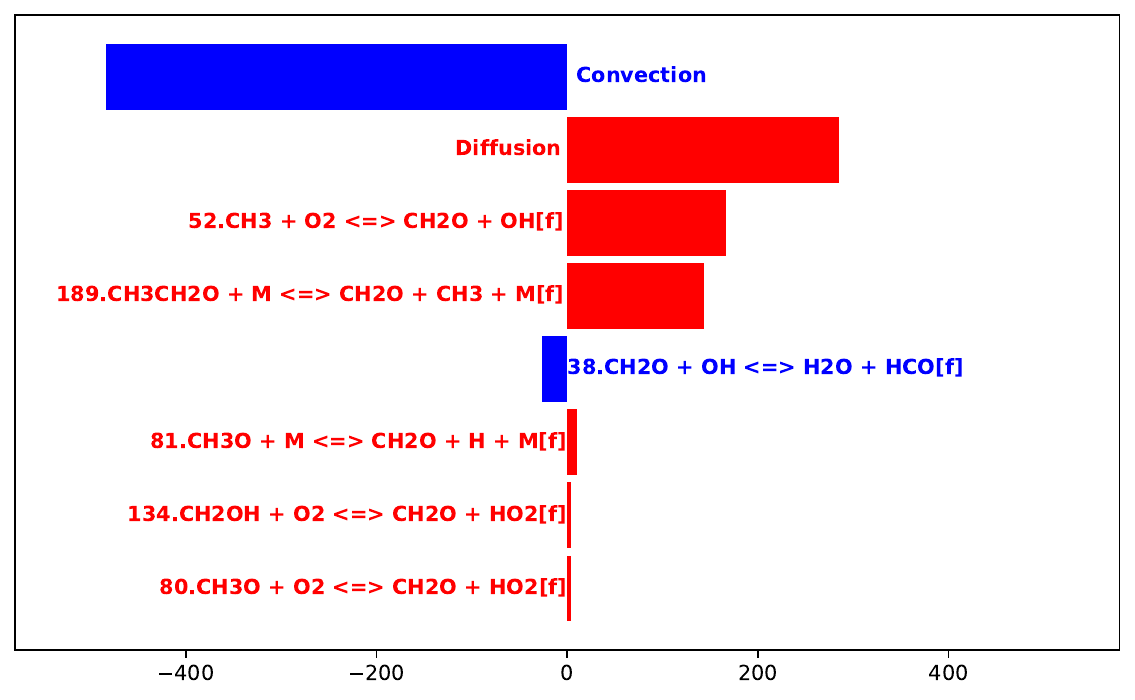}}
\captionsetup{ width=\textwidth}
\caption{Contribution on {\chto} concentration change @4.2mm}
\label{fig:CH2O}
\end{figure}

\begin {figure} [H]
\centerline{\includegraphics[width = 192pt]{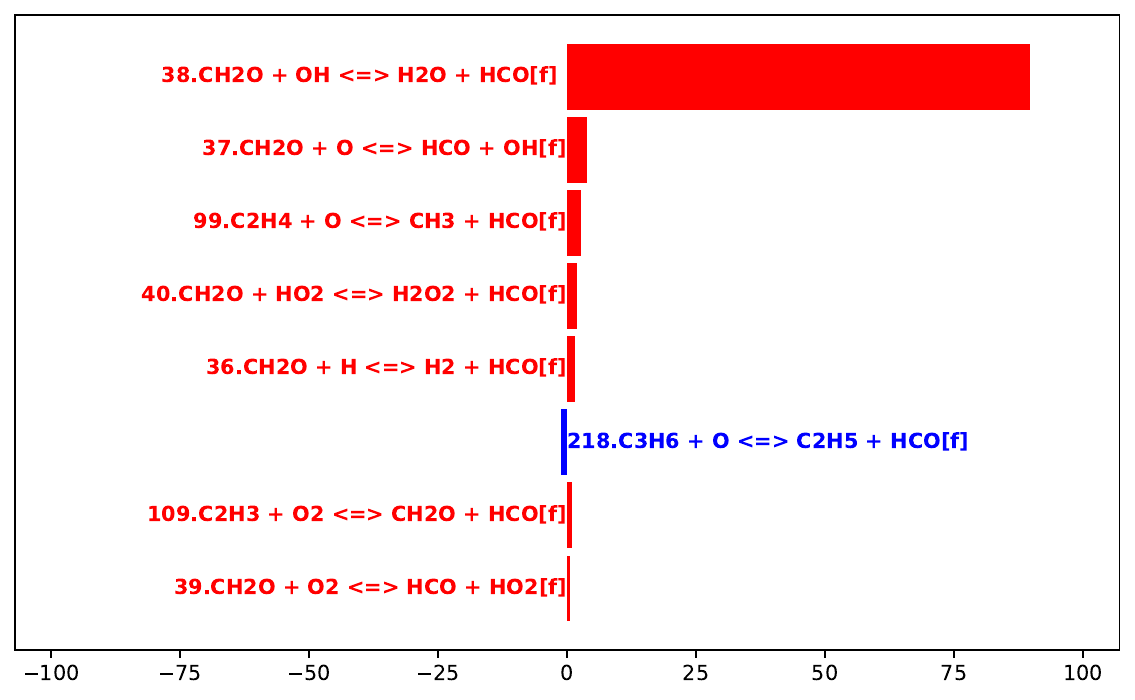}}
\captionsetup{ width=\textwidth}
\caption{Contribution on {\hco} concentration change @4.2mm}
\label{fig:HCO}
\end{figure}

\begin {figure} [H]
\centerline{\includegraphics[width = 192pt]{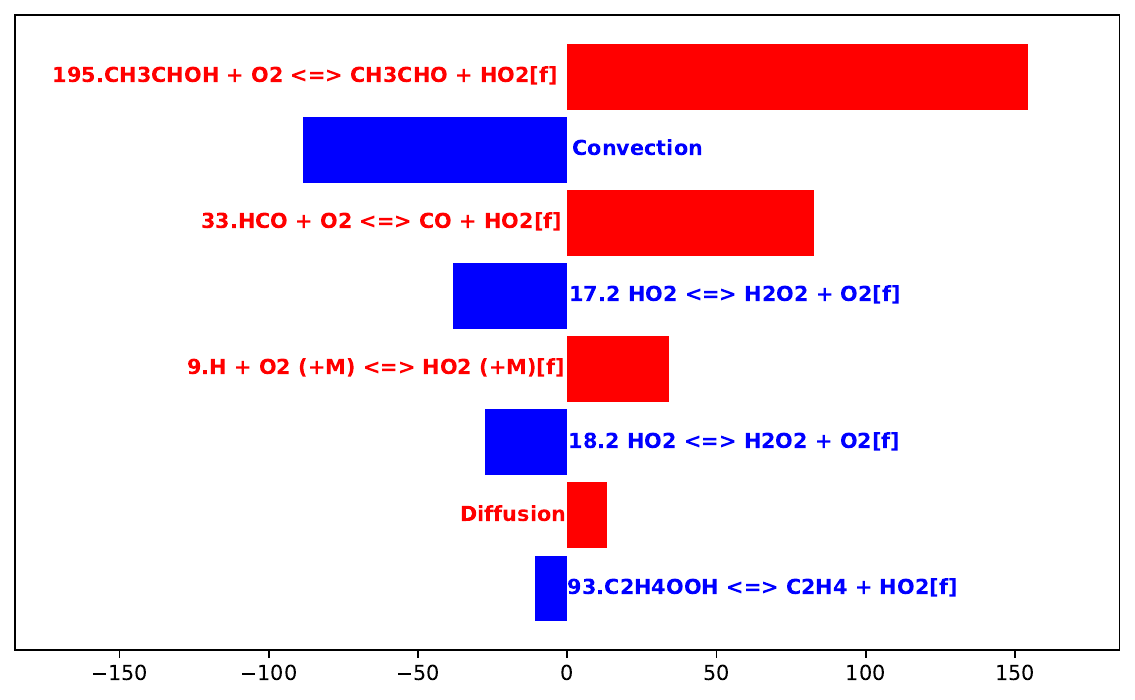}}
\captionsetup{ width=\textwidth}
\caption{Contribution on {\hot} concentration change @4.2mm}
\label{fig:HO2}
\end{figure}

\section{Details of Derivation of Analysis Method}
\subsection{Derivation of reaction-rate-change-analysis}
\label{Appendix:b1}
The first-order derivatives, ${\partial \dot {\omega}_n}/{\partial k_n}$ and ${\partial \dot {\omega}_n}/{\partial c_i}$, correspond to the changes caused by variations in individual variables. However, the reaction rate expression, ${\dot{\omega}_n}$, is a multivariate function. For instance, $\dot{\omega}_{f,k} = k_{f,k} \prod_{i=1}^{m} c_i^{\nu'i}$ and $\dot{\omega}_{b,k} = k_{b,k} \prod_{i=1}^{m} c_i^{\nu''_i}$. Therefore, to enhance precision, it is essential to consider the reaction rate changes resulting from the mutual influence of variables. For this purpose, the second-order Taylor expansion can be employed, offering a more comprehensive understanding and accurate calculation. 

The formula below outlines the approximated reaction rate from the second-order Taylor expansion for function of multiple variables:
\begin {equation}
{\Delta \dot{\omega}^\text{approx}_n= \displaystyle \sum_{i=0}^{m} \frac{\partial \dot{\omega}_n}{\partial x_i} \Delta {x_i} + \frac{1}{2} \sum_{i=0}^{m} \sum_{j=0}^{m} \frac{\partial^2 \dot{\omega}_n}{\partial x_i \partial x_j} \Delta x_i \Delta x_j} 
\end{equation}

For simplification, let the variable ${x_i}$ encompasses both the rate constant, $k_n$, and the species concentration, $c_i$, with ${x_0}$ specifically denoting ${k_n}$.
\begin{equation}
  x_i  = 
    \begin{cases} \displaystyle {k_n}, i = 0\\
    \displaystyle c_i, i \neq 0
    \end{cases} \text{where} i =0,1 ...
\end{equation}

By employing the second derivative in Taylor series, the change in reaction rate attributed to $\Delta x_i$ could be written as:
\begin{equation}
{\Delta \dot{\omega}_n|_{x_i}= \frac{\Delta \dot {\omega}_n}{\Delta \dot{\omega}^{\text{approx}}_n} \times (\displaystyle \frac{\partial \dot{\omega}_n}{\partial x_i} \Delta {x_i} + \frac{1}{2} \frac{\partial^2 \dot{\omega}_n}{\partial x_i^2} (\Delta x_i)^2+ \sum_{\substack{j=0\\j \neq i}}^{m} \frac{1}{2} \frac{\partial^2 \dot{\omega}_n}{\partial x_i \partial x_j} \Delta x_i \Delta x_j)} 
\end{equation}

The term $\frac{\partial \dot{\omega}_n}{\partial x_i} \Delta {x_i}+\frac{1}{2} \frac{\partial^2 \dot{\omega}_n}{\partial x_i^2} (\Delta x_i)^2$ delineates the direct contribution of $\Delta {x_i}$ to $\Delta \dot{\omega}_n$, encapsulating both linear and quadratic influences. While, the term $\frac{1}{2} \frac{\partial^2 \dot{\omega}_n}{\partial x_i \partial x_j} \Delta x_i \Delta x_j$ quantifies the synergistic influence on $\Delta \dot{\omega}_n$ emanating from interaction between $\Delta {x_i}$ and $\Delta {x_j}$. This synergistic influence could be symmetrically attributed to the influence from both $\Delta {x_i}$ and $\Delta {x_j}$.

\subsection{Derivation of species concentration change}
\label{Appendix:b2}

For the counterflow flame, in steady state solution, governing equation for species $i$ is given by :

\begin{equation}
\displaystyle
0=-{\rho}u\frac{dY_i}{dz}-\frac{d {j_i}}{dz}+{W_i  (\dot{\omega}_i^+ -  \dot{\omega}_i^-})
\label{app_eq:gvn_eq}
\end{equation}

Here, $W_i$ denotes the molecular weight of species $i$. The production rate of species $i$ is represented by $\dot{\omega}_i^+$, while $ \dot{\omega}_i^-$ symbolizes the consumption rate of species $i$. The production and consumption rate can be expressed as follows:

\begin{align}
\dot{\omega}_i^+&=  \sum_{k}\nu_{i,k} \left[(1-\delta'_{i,k})  \dot{\omega}_{f,k} - \delta'_{i,k}  \dot{\omega}_{b,k} \right] \nonumber\\
&=  \sum_{k} \lvert \nu_{i,k} \rvert \left[(1-\delta'_{i,k})  \dot{\omega}_{f,k} + \delta'_{i,k}  \dot{\omega}_{b,k} \right]
\label{app_eq:prod}
\end{align}

\begin{align}
\dot{\omega}_i^- &=  \sum_{k} \nu_{i,k} \left[ -\delta'_{i,k}  \dot{\omega}_{f,k} + (1-\delta'_{i,k})  \dot{\omega}_{b,k} \right]\nonumber\\
&=  \sum_{k} \lvert \nu_{i,k} \rvert \left[ \delta'_{i,k}  \dot{\omega}_{f,k} + (1-\delta'_{i,k})  \dot{\omega}_{b,k} \right]
\label{app_eq:consm}
\end{align}

with $\delta'_{i,k}$ defined as
\begin{equation}
\delta'_{i,k}  =
\begin{cases}
1 & \text{if } \nu_{i,k}  < 0, \\
0 &  \text{if } \nu_{i,k} > 0
\end{cases}
\end{equation}

The term \[ \nu_{i,k} = \nu''_{i,k} - \nu'_{i,k} \] 
where $\nu'_{i,k}$, $\nu''_{i,k}$ are the stoichiometric coefficients for species $i$ appearing as a reactant and as a product in $k^{th}$ reaction, respectively.

Governing equation of species $i$, Eqn.\eqref{app_eq:gvn_eq}, could be rewritten as :
\begin{equation}
\displaystyle
\dot{\omega}_i^-=-{\rho}u\frac{dY_i}{dz} /W_i -\frac{d {j_i}}{dz} /W_i + \dot{\omega}_i^+
\label{app_eq:sp_eq2}
\end{equation}

While, $\dot{\omega}_i^-$ can be expressed as:
\begin{equation}
\displaystyle
\dot{\omega}_i^- = c_i \times {\sum_{k}\lvert\nu_{i,k}\rvert \left[\delta'_{i,k}  w^{den}_{f,k}+(1-\delta'_{i,k})w^{den}_{b,k}\right]}
\label{app_eq:omegaB-}
\end{equation}
where $  w^{den}_{f,k},  w^{den}_{b,k} $ are defined as:

 \begin{equation}
 \displaystyle w^{den}_{f,k}  =k_{f,k} \prod_{j=1}^{n} c_i^{\nu'_j-\delta_{i,j} }
 ,w^{den}_{b,k}  = k_{b,k}\prod_{j=1}^{n} c_i^{\nu''_j-\delta_{i,j} }
\end{equation}

 and $ \delta_{i,j}$ indicates whether species $j$ is species $i$:
 \begin{equation}
 \delta_{i,j}  =
\begin{cases}
1 & \text{if the species $j$ is species $i$}, \\
0 & \text{otherwise}.
\end{cases}
\end{equation}

Combining the Eqn.\ \eqref{app_eq:sp_eq2} and Eqn.\ \eqref{app_eq:omegaB-}, the concentration of substance $i$, $ c_i$ can be derived from following expression:

\begin{equation}
 c_i = \frac{\displaystyle  -{{\rho}u\frac{dY_i}{dz}}/{W_i} -{\frac{d {j_i}}{dz}}/{W_i} + \dot{\omega}_i^+} {\displaystyle\sum_{k} \lvert\nu_{i,k}\rvert \left[\delta'_{i,k} w^{den}_{f,k}+(1-\delta'_{i,k})w^{den}_{b,k}\right]}
\end{equation}

Upon substituting Eqn.\ \eqref{app_eq:prod} into the equation above, we could obtain a detailed expression for $ c_i$ as:
\begin{equation}
 c_i =\frac{\displaystyle -{{\rho}u\frac{dY_i}{dz}}/{W_i} -{\frac{d {j_i}}{dz}}/{W_i}+ \sum_{k} \lvert \nu_{i,k} \rvert \left[(1-\delta'_{i,k})  \dot{\omega}_{f,k} + \delta'_{i,k}  \dot{\omega}_{b,k} \right]}  {\displaystyle\sum_{k} \lvert\nu_{i,k}\rvert \left[\delta'_{i,k} w^{den}_{f,k}+(1-\delta'_{i,k})w^{den}_{b,k}\right]}
\label{app_eq:xb2}
\end{equation}


 \footnotesize
 \baselineskip 9pt


\bibliographystyle{elsarticle.bst}
\bibliography{reference}


\newpage

\small
\baselineskip 10pt



\end{document}